\documentclass[a4paper,11pt]{article}
\pdfoutput=1

\usepackage{jheppub} 

\usepackage{amsmath,amssymb,graphicx,multirow,xspace}
\usepackage[colorlinks=true,urlcolor=blue,anchorcolor=blue,citecolor=blue,filecolor=blue,linkcolor=blue,menucolor=blue,pagecolor=blue]{hyperref}
\usepackage{subfigure, placeins}

\long\def\symbolfootnote[#1]#2{\begingroup%
\def\thefootnote{\fnsymbol{footnote}}\footnote[#1]{#2}\endgroup}

\newcommand{\newc}{\newcommand}
\newc{\gsim}{\lower.7ex\hbox{$\;\stackrel{\textstyle>}{\sim}\;$}}
\newc{\lsim}{\lower.7ex\hbox{$\;\stackrel{\textstyle<}{\sim}\;$}}
\newc{\gev}{\,{\rm GeV}}
\newc{\mev}{\,{\rm MeV}}
\newc{\ev}{\,{\rm eV}}
\newc{\kev}{\,{\rm keV}}
\newc{\tev}{\,{\rm TeV}}

\newc{\mz}{M_Z}
\newc{\mpl}{M_*}
\newc{\mw}{m_{\rm weak}}
\newc{\nr}[1]{N^c_R{}_{#1}}
\usepackage{amsmath}

\renewcommand{\dag}{\dagger}

\newcommand{\Q}{\bar Q}
\renewcommand{\O}{O}

\newcommand{\mhu}{{\hat m_{H_u}}}
\newcommand{\mhd}{{\hat m_{H_d}}}
\newcommand{\mhud}{{\hat m_{H_{u,d}}}}

\def\beq{\begin{equation}}
\def\eeq{\end{equation}}
\def\bea{\begin{eqnarray}}
\def\eea{\end{eqnarray}}
\def\bitem{\begin{itemize}}
\def\eitem{\end{itemize}}
\newc{\ie}{{\it i.e.}}          \newc{\etal}{{\it et al.}}
\newc{\eg}{{\it e.g.}}          \newc{\etc}{{\it etc.}}
\newc{\cf}{{\it c.f.}}

 \numberwithin{equation}{section}

\newcommand\fverb{\setbox\fverbbox=\hbox\bgroup\verb}
\newcommand\fverbdo{\egroup\medskip\noindent%
            \fbox{\unhbox\fverbbox}\ }
\newcommand\fverbit{\egroup\item[\fbox{\unhbox\fverbbox}]}
\newbox\fverbbox

\begin{document}

\begin{titlepage}

\begin{flushright}
\small{RUNHETC-2013-25}\\
\end{flushright}

\vspace{0.5cm}
\begin{center}
\Large\bf

Higgs Mediation with \\  Strong Hidden Sector Dynamics

\end{center}

\vspace{0.2cm}
\begin{center}
{\sc Simon Knapen
and David Shih}\\
\vspace{0.6cm}
\textit{New High Energy Theory Center\\
Department of Physics and Astronomy\\
Rutgers University, Piscataway, NJ 08854, USA}\\
\end{center}

\vspace{0.4cm}

\begin{abstract}

We present a simple model that achieves $m_h\approx 126$ GeV in the MSSM with large $A$-terms and TeV-scale stops through a combination of gauge mediation and Higgs-messenger interactions. The $\mu$/$B_\mu$ and $A$/$m_H^2$ problems are both solved by a common mechanism -- partial sequestering from strong hidden sector dynamics. Using the framework of General Messenger Higgs Mediation, we explicitly calculate the soft masses in terms of the vacuum expectation values, operator dimensions and OPE coefficients of the strongly-coupled hidden sector.  Along the way, we also present a general analysis of the various constraints on sequestered Higgs mediation models. The phenomenology of such models is similar to gaugino mediation, but with large $A$-terms. The NLSP is always long-lived and is either the lightest stau or the Higgsino.  The colored states are typically out of reach of the 8 TeV LHC, but may be accessible at 14 TeV, especially if the NLSP is the lightest stau.
\end{abstract}

\end{titlepage}

\vspace{0.2cm}
\noindent

\tableofcontents

\section{Introduction}

The discovery of a Higgs boson near 126 GeV \cite{Aad:2012tfa,Chatrchyan:2012ufa} has profound implications for supersymmetry as a solution to the electroweak hierarchy problem. This is especially the case in minimal supersymmetry, where the stops must either be unnaturally heavy
($\gtrsim 10$ TeV) or have a large trilinear coupling to the Higgs \cite{Hall:2011aa, Heinemeyer:2011aa, Arbey:2011ab, Arbey:2011aa, Draper:2011aa, Carena:2011aa, Cao:2012fz,  Christensen:2012ei, Brummer:2012ns}. The former possibility leaves little hope for preserving naturalness or observational signals at the LHC, so we will focus on the latter scenario.
This requires a plausible mechanism for generating such large $A$-terms without introducing large flavor violation or other unwanted effects.   

The lack of decisive deviations in searches for flavor and CP violation has long favored low-scale gauge mediation by virtue of its flavor universality. However, in its minimal form, gauge mediation is challenged by the Higgs sector, since it generates neither the $\mu$ and $B_\mu$ parameters necessary for electroweak symmetry breaking (EWSB), nor the $A$-terms suggested by the Higgs mass measurement. These terms may be generated in a flavor universal manner by adding interactions between the Higgs sector and the messenger sector, 
\begin{equation} \label{eq:Wud}
W \supset \lambda_u{O}_u H_u + \lambda_d{O}_d H_d
\end{equation}
where ${O}_{u,d}$ are messenger-sector operators. 
 Although the $\mu$ and $A$-terms are obtained trivially in such a setup, viable solutions must confront two thorny problems: the ``$\mu$/$B_\mu$ problem'' \cite{Dvali:1996cu} and the ``$A$/$m_H^2$ problem'' \cite{Craig:2012xp}. Both problems arise because adding Higgs-messenger interactions that generate a $\mu$ ($A$) term also tend to produce a $B_\mu$ ($m_H^2$) term that is too large for viable electroweak symmetry breaking.  
 
 The most stringent form of the $A$/$m_H^2$ problem may be resolved if the sole source of messenger mass is a single SUSY-breaking spurion \cite{Giudice:2007ca,Craig:2012xp}, as in minimal gauge mediation (MGM) \cite{Dine:1993yw,Dine:1994vc,Dine:1995ag}. But even in this case the $\mu$/$B_\mu$ problem remains unaddressed, and requires a further extension of the model. Moreover, there is a residual ``little  $A$/$m_H^2$ problem'', as any weakly-coupled model that generates large $A$-terms through Higgs-messenger interactions also generates contributions to the Higgs soft masses proportional to $A^2$ \cite{Craig:2012xp}. Even if these contributions do not prevent  electroweak symmetry breaking, they significantly increase the fine-tuning associated with the weak scale.   

In this paper, we present an alternative framework which uses strong dynamics in the hidden sector to economically solve both the $\mu$/$B_\mu$ and $A$/$m_H^2$ problems. 
Two ingredients are required for this: that there exists a hierarchy between the messenger scale $M$ and the SUSY-breaking scale $\sqrt{F}$; and that the anomalous dimensions of the operators responsible for SUSY-breaking are large and positive. The former property is a generic prediction of dynamical supersymmetry breaking \cite{Witten:1981nf}, while the latter property is constrained, but still allowed by the conformal bootstrap \cite{Poland:2011ey}. If both these conditions are met, strong renormalization effects in the hidden sector can suppress the soft masses of the scalars (including $B_\mu$ \cite{Dine:2004dv, Roy:2007nz, Murayama:2007ge} and $m_H^2$ \cite{Craig:2013wga}), an idea more generally known as ``conformal sequestering'' or ``scalar sequestering'' \cite{Randall:1998uk,Luty:2001jh,Luty:2001zv}. We will demonstrate that with such a strongly coupled hidden sector, even the very simplest example for the messenger sector yields a large viable parameter space. The simplicity of our model contrasts sharply with most fully weakly-coupled solutions, which address the $\mu$/$B_\mu$ problem by elaborately extending either the Higgs sector or the messenger sector (or both).

In recent years, there has been tremendous progress in our understanding of 4D conformal field theory, starting with the work of \cite{Rattazzi:2008pe}. This revival of the conformal bootstrap program has led to strong bounds on the dimensions of operators appearing in the OPE. Applying these bounds to the operators responsible for SUSY-breaking has in turn strongly limited the efficacy of the conformal sequestering scenario \cite{Poland:2011ey}. In particular, it is now very difficult to achieve full suppression of $B_\mu$ and $m_{H_u}^2$,
\begin{align}
B_\mu\ll |\mu|^2 \quad\mathrm{and}\quad m_{H_u}^2+|\mu|^2 \ll |A_u|^2.
\end{align}
On the other hand, a \emph{partial} suppression of the dangerous contributions such that 
\begin{align}
B_\mu\lesssim |\mu|^2 \quad\mathrm{and}\quad m_{H_u}^2+|\mu|^2 \lesssim |A_u|^2.
\end{align}
is still possible and may be sufficient to facilitate electroweak symmetry breaking. In this case the details of the hidden sector dynamics do not fully decouple from the low energy observables, and testing for viable electroweak symmetry breaking requires a robust framework to explicitly compute the MSSM soft parameters in terms of the hidden sector data (such as the spectrum of operators, their scaling dimensions, and their OPE coefficients). 

\begin{figure}\centering
\includegraphics[width=0.85\textwidth]{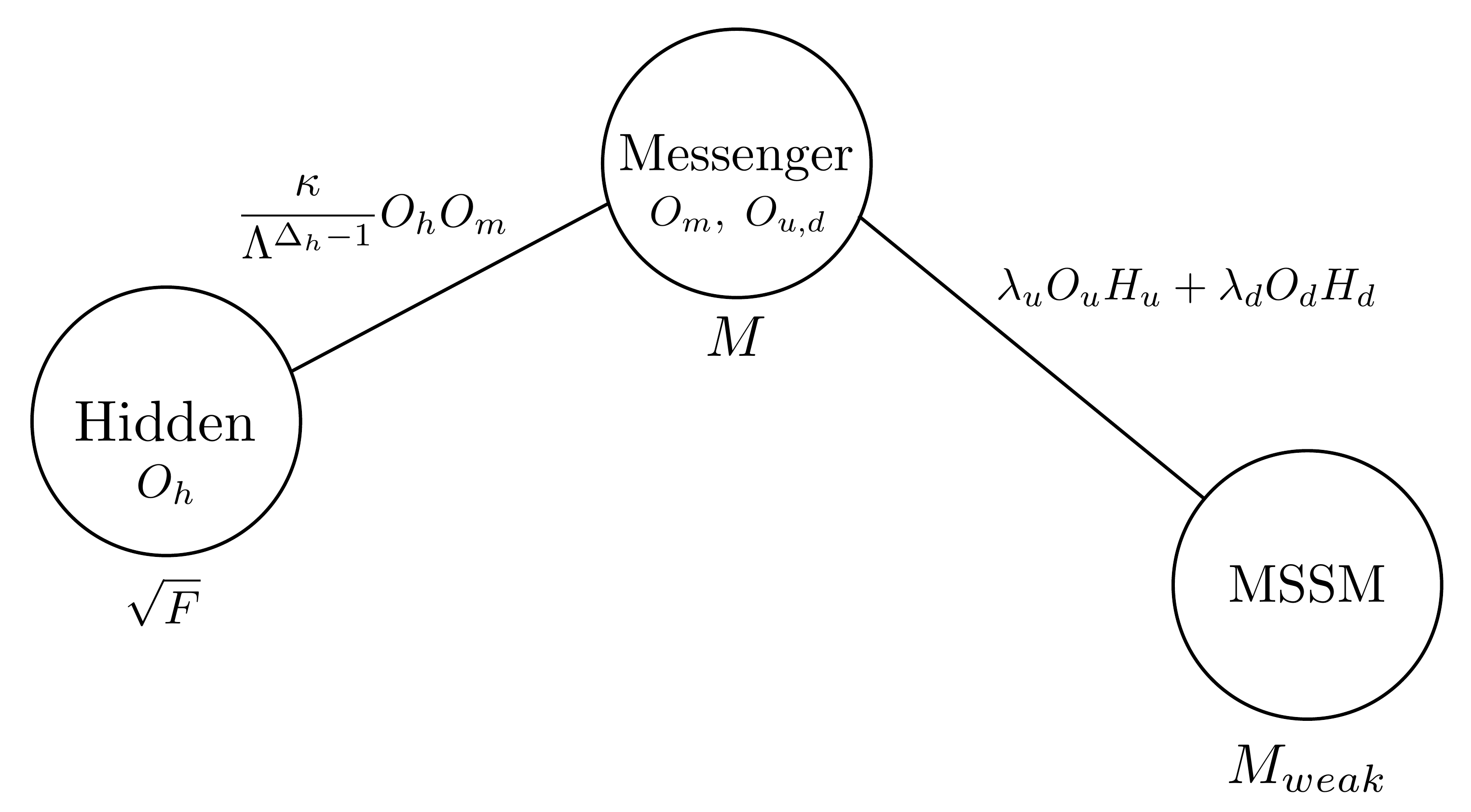}
\caption{Schematic representation of the various sectors and couplings. This paper we take the messenger sector to be weakly coupled but allow for strong dynamics in the hidden sector.\label{GMHMdiagram}}
\end{figure}

General Messenger Higgs Mediation (GMHM), developed recently in \cite{Craig:2013wga}, provides precisely such a framework. Following \cite{Dumitrescu:2010ha}, the idea of GMHM is to go beyond the single-sector frameworks of \cite{Meade:2008wd,Buican:2008ws,Komargodski:2008ax} and explicitly separate the messenger sector and SUSY-breaking sector, so that it becomes possible to take $\sqrt{F}\ll M$. Specifically, we parametrize the coupling between the messenger sector and the SUSY-breaking hidden sector via a  perturbative superpotential interaction as in  \cite{Dumitrescu:2010ha}:
\begin{equation} \label{eq:Whm}
W \supset \frac{\kappa}{\Lambda^{\Delta_h-1}}{O}_h{O}_m
\end{equation}
where ${O}_h$ is an operator in the SUSY-breaking sector with dimension $\Delta_h$, ${O}_m$ is an operator in the messenger sector, and $\Lambda$ is the cut-off scale associated with the irrelevant operator in (\ref{eq:Whm}). The complete setup of GMHM is shown in figure  \ref{GMHMdiagram}. 
By expanding in the portal couplings $\kappa$, $\lambda_{u,d}$ of (\ref{eq:Wud}) and (\ref{eq:Whm}), we can express the soft parameters in terms of products of separate correlation functions over the messenger sector and the hidden sector. Under the assumption that the hidden sector is near a conformal fixed point between the scales $M$ and $\sqrt{F}$, the correlators simplify dramatically. The GMHM formalism then allows, for the first time, for a full calculation of soft masses directly in terms of hidden sector scaling dimensions, OPE coefficients, and expectation values. 

Although GMHM applies to any hidden sector and messenger sector coupled through the portals (\ref{eq:Wud}), (\ref{eq:Whm}), in this paper we will focus on weakly-coupled messenger sectors in order to preserve calculability and predictivity.\footnote{For this reason we will  assume for simplicity that $\Delta_{m}=\Delta_{u}=\Delta_{d}=2$ (while allowing for arbitrary $\Delta_h$), which is well motivated for a weakly coupled messenger sector. This explains the powers of $\Lambda$ or lack thereof implicitly taken in (\ref{eq:Wud}) and (\ref{eq:Whm}).}
We will explore the phenomenology of this entire class of models, as well as present a very simple explicit example.  Concretely, the model for the messenger sector that we consider is given by
\begin{align}
W=\left(\frac{\kappa}{\Lambda^{\Delta_h-1}}{O}_h+M\right)\left(\tilde\phi_D\phi_D+ \tilde\phi_S\phi_S\right)+\lambda_{u}\tilde\phi_D\phi_S H_u+\lambda_{d} \phi_D\tilde\phi_S H_d.\label{Wproxy}
\end{align}
where $\phi_D$, $\tilde\phi_D$ and $\phi_S$, $\tilde\phi_S$ are $SU(2)$ doublets and gauge singlets respectively. Although this model is the prime example of a model that does {\it not} solve the $\mu/B_\mu$ problem when the hidden sector is trivial  \cite{Dvali:1996cu}, with partial hidden-sector sequestering it becomes an elegant solution to both the $\mu/B_\mu$ and $A/m_H^2$ problems. We find that electroweak symmetry breaking and $m_h=126$~GeV are easy to achieve in this model, and instead the most interesting constraints on the parameter space originate from  stau tachyons. Nevertheless there is a large viable parameter space, which can accommodate ${\mathcal O}(1)$ OPE coefficients and roughly 10\% suppression from conformal sequestering. The collider phenomenology is similar to that of standard gaugino mediation \cite{Chacko:1999mi,Kaplan:1999ac}, with all the colored states above 1~TeV. The NLSP is always long-lived, which leads to spectacular collider signatures if the NLSP is a stau.

The paper is organized as follows. Section 2 is a brief review of the mechanism of conformal sequestering as well as the most important features of the GMHM formalism. In section 3 we discuss the model-independent constraints on the parameter space from weak scale requirements such as EWSB and the Higgs mass, prior to presenting a full analysis of our explicit example in section 4. Section 5 is a short discussion of the collider phenomenology of this class of models. Section 6 contains our conclusions, and we reserve various technical details for the appendices.

\section{Review of GMHM and Conformal Sequestering \label{secreview}}

\subsection{The GMHM formalism}

In this subsection, we review the calculation of the Higgs soft parameters $\mu$, $B_\mu$, $A_{u,d}$ and $m_{H_{u,d}}^2$ through the GMHM formalism. For the derivation of the various results   we refer to \cite{Craig:2013wga}. At the scale $\sqrt{F}$, conformal symmetry and supersymmetry are broken by an $F$-term expectation value for the hidden sector operator $O_h$ with dimension $\Delta_h$: 
\begin{align}\label{onepointvev}
\langle Q^2 \O_h \rangle_h \equiv \sqrt{F}^{\Delta_h + 1},
\end{align}
To leading order, the dimension-one soft parameters (gaugino masses, $\mu$, and $A_{u,d}$) are only sensitive to this vacuum expectation value.  

Meanwhile, the dimension-two soft parameters (sfermion mass-squareds, $B_\mu$, and $m_{H_{u,d}}^2$) are sensitive to the dynamics of the hidden sector. The leading contribution of such dynamics is packaged in the hidden-sector two-point function 
\begin{align}\label{hidsectwopoint}
\langle Q^4 [\O_h^\dag(x) \O_h(x')] \rangle_h,
\end{align}
In the spurion limit, this correlation function simply factorizes into $|\langle Q^2 \O_h \rangle_h|^2$, but in a non-trivial hidden sector this is not necessarily a good approximation. For calculable models of the supersymmetry breaking sector one could address this issue by explicitly evaluating (\ref{hidsectwopoint}) and then studying its effects on the low energy physics. In this paper we will take a different approach: we will remain agnostic about the precise mechanism of supersymmetry breaking, but instead assume that the hidden sector is approximately conformal before it breaks SUSY. In the GMHM framework, the hidden sector correlator in (\ref{hidsectwopoint}) is always convolved with a short-distance messenger correlator, which then enforces $|x-x'|\sim\frac{1}{M}\ll \frac{1}{\sqrt{F}}$. It is therefore justified to simplify (\ref{hidsectwopoint}) by making use of the operator product expansion:
\begin{equation}
\O_h(x) \O_h^\dag(x') \sim |x-x'|^{-2 \Delta_h}\mathbf{1} + C |x - x'|^{\Delta - 2 \Delta_h} \O_{\Delta}(x) + \dots
\end{equation}
where the ellipses denote terms with higher dimension and/or spin. The supercharges annihilate the unit operator such that the correlation function is reduced to
\begin{equation}\label{hidsectwopointsimple}
\langle Q^4 [\O_h(x) \O_h^\dag(x')]\rangle_h \approx C |x - x'|^{\Delta - 2 \Delta_h} \langle Q^4\O_{\Delta}\rangle_h
\end{equation}
where we only keep the leading non-vanishing term in the OPE.  Dimensional analysis then demands that the $D$-term expectation value of $\O_\Delta$ takes the form $\langle Q^4 \O_\Delta \rangle \equiv \xi_{\Delta} F^{(\Delta + 2)/2}$, where $\Delta$ is the scaling dimension of $\O_\Delta$ and $\xi_\Delta$ is a dimensionless number. The parameters $\xi_\Delta$ and $C$ are degenerate at the level of our analysis, and to facilitate the notation we thus introduce an `effective OPE coefficient' :
\begin{equation}
\hat C \equiv C \xi_{\Delta}.
\end{equation}
Note that $\xi_\Delta$ (and therefore $\hat C$) is a real number, but can have either sign.

To leading order in $\lambda_{u,d}$, the $\mu$ term and the $A$-terms are given by
\begin{align} \label{eq:mu}
\mu &= -\lambda_u \lambda_d \kappa\frac{\sqrt{F}^{\Delta_h+1}}{\Lambda^{\Delta_h-1}}  \int d^4 y \, d^4 x \langle \O_m^\dag(y) Q^\alpha \O_u(x) Q_\alpha \O_d(0) \rangle_{m}  \\
A_{u,d} &= |\lambda_{u,d}|^2 \kappa \frac{\sqrt{F}^{\Delta_h+1}}{\Lambda^{\Delta_h-1}} \int d^4 y d^4 x \langle  \O_m^\dagger (y) \Q^2\big[\O_{u,d}^\dag(x) \O_{u,d}(0) \big]\rangle_m  \label{eq:A}
\end{align}
The Higgs sector soft masses are specified by the correlators
\begin{align}
B_\mu &= - \lambda_u\lambda_d \kappa^2  \hat C  \frac{\sqrt{F}^{\Delta+2}}{\Lambda^{2\Delta_h-2}} \int d^4y\,d^4y'\,d^4x\,  |y-y'|^{\gamma} 
  \Big\langle \O_m(y)\O_m^\dagger(y')Q^2 \O_u(x)Q^2 \O_d(0)\Big\rangle_{m}\label{bmucor}\\
  \mhud^2 &=- |\lambda_{u,d}|^2 \kappa^2 \hat C  \frac{\sqrt{F}^{\Delta+2}}{\Lambda^{2\Delta_h-2}} \int d^4y\,d^4y'\,d^4x\,   |y-y'|^{\gamma}  \Big\langle \O_m(y)\O_m^\dagger(y')Q^2 \O_{u,d}(x) \Q^2 \O^\dagger_{u,d}(0)\Big]\Big\rangle_{m}\label{mhcor}
\end{align}
with $\gamma\equiv\Delta-2\Delta_h$. Here we have introduced the following notational convenience:
\begin{align}
\mhud^2\equiv m^2_{H_{u,d}}+|\mu|^2
\end{align}
where the $m^2_{H_{u,d}}$ are the usual soft masses for the Higgs fields.

Although the Higgs sector parameters are generated by the portal (\ref{eq:Wud}), for the rest of the MSSM soft parameters we need a different source. In this paper, we assume that these arise through standard gauge mediation, i.e.\ the messenger sector in figure \ref{GMHMdiagram} also couples to the MSSM through gauge interactions. For completeness, let us exhibit the usual gauge-mediated contributions to the soft masses. These can be assembled from the GGM correlators \mbox{\cite{Meade:2008wd,Buican:2008ws}:}
\begin{align}
M_i&=g_i^2 B_i\nonumber\\
m_{\tilde f}^2 &=\sum^3_{i=1}g^4_i c_2(f,i) A_i\label{GGMforms}
\end{align}
where $f$ labels the matter representations of the MSSM, and $c_2(f, i)$ is the quadratic Casimir of $f$ with respect to the gauge group $i$. In the GMGM formalism the $B_i$ and $A_i$ correlators can be written as a convolution of a messenger sector correlator with a hidden sector correlator \cite{Dumitrescu:2010ha}. Crucially, the hidden sector correlator appearing in the expression for the $A_i$ is precisely (\ref{hidsectwopoint}). Using the OPE, the expressions for $B_i$ and $A_i$ then reduce to 
\begin{align}
B_i &=  \frac{\kappa}{4}\frac{\sqrt{F}^{\Delta_h+1}}{\Lambda^{\Delta_h-1}}  \int d^4 y \, d^4 x \langle Q^2 O_m^\dag(y) J_i(x) J_i(0) \rangle_{m}\label{Bcor}\\
A_i &=-\frac{\kappa^2}{128\pi^2}  \hat C  \frac{\sqrt{F}^{\Delta+2}}{\Lambda^{2\Delta_h-2}} \int d^4y\,d^4y'\,d^4x\,  |y-y'|^{\gamma}
  \Big\langle Q^4\big[ \O_m(y)\O_m^\dagger(y')\big]J_i(x) J_i(0)\Big\rangle_{m} \log[M^2 x^2]\label{Acor}
\end{align}
The $J_i(x)$ are the bottom components of the current superfields through which the messengers couple to gauge group $i$. In contrast with (\ref{bmucor}) and (\ref{mhcor}), the expression for sfermion mass-squareds in (\ref{Acor}) is suppressed by an extra loop factor, in addition to any loop factors that may be generated by the messenger correlator itself.

\subsection{Conformal sequestering}
\label{subsec:conformalsequestering}

For a generic weakly-coupled messenger sector, all the messenger correlators in equations (\ref{eq:mu})-(\ref{mhcor}) are non-zero at one loop, which implies that $B_\mu$ and $\mhud^2$ are too large to facilitate viable electroweak symmetry breaking. However just by applying naive dimensional analysis on the correlators in the previous section, we can already identify several possible avenues to address the problem: 
\begin{align}
\frac{B_\mu}{\mu^2}&\sim\frac{16\pi^2}{\lambda_u\lambda_d}\frac{\hat C}{N} \left(\frac{\sqrt{F}}{M}\right)^{\gamma}\nonumber\\
\frac{\mhud^2}{|A_{u,d}|^2}&\sim\frac{16\pi^2}{|\lambda_{u,d}|^2}\frac{\hat C}{N} \left(\frac{\sqrt{F}}{M}\right)^{\gamma}\label{dimanmf}
\end{align}
A well known method to mitigate the infamous loop factor is to increase the messenger number, which we denote by $N$. However, this is limited by Landau poles in the gauge couplings and cannot be responsible for completely suppressing the loop factor. Secondly, if $\gamma>0$ and $\sqrt{F}\ll M$, the last factor on each line of (\ref{dimanmf}) can in principle suppress the loop factor.  This is the conformal sequestering mechanism. Finally, one could consider an SCFT with $\hat C\ll 1$, such that the effective OPE coefficient provides the desired suppression factor, possibly in combination with some suppression from sequestering. 

Meanwhile, from equations (\ref{GGMforms})-(\ref{Acor}) we see that since the gaugino and sfermion masses are generated through gauge mediation, they satisfy:
\begin{align}
\frac{m^2_{\tilde f}}{M_i^2}&\sim\frac{\hat C}{N} \left(\frac{\sqrt{F}}{M}\right)^{\gamma} \label{eq:gauginosfermion}
\end{align}
In particular, the sfermion masses come with an extra loop factor with respect to $B_\mu$ and $\mhud^2$, but are subject to the same suppression from the hidden sector. This implies that the sfermion masses are always suppressed with respect to the gaugino masses if the $\mu$/$B_\mu$ and $A$/$m_H^2$ problems are solved. The phenomenology will therefore be similar to that of gaugino mediation \cite{Chacko:1999mi,Kaplan:1999ac}.

The idealized cases where $\gamma\gg1$ or $\hat C\rightarrow 0$ lead to the extremely simple boundary conditions at the scale $\sqrt{F}$:
\begin{align}
B_\mu \approx  \mhud^2\approx  m^2_{\tilde f}\approx 0\label{fullyseq}
\end{align}
Interestingly, this part of the UV boundary conditions becomes completely model-independent. The sensitivity of these parameters to the details of the hidden sector and messenger sector has been completely erased.

Unfortunately this scenario is severely challenged in several ways. First, it has been known for some time that achieving suitable EWSB is nontrivial for these boundary conditions \cite{Perez:2008ng,Asano:2008qc}. Second, even if one succeeds in breaking electroweak symmetry, the amount of sequestering through the factor $\Big(\frac{\sqrt{F}}{ M}\Big)^{\gamma}$ is now severely limited by powerful upper bounds on $\gamma$ from the internal consistency of the hidden sector SCFT \cite{Poland:2011ey}.

\begin{figure}[t!]\centering
\subfigure[$\epsilon=1$]{\includegraphics[width=0.45\textwidth]{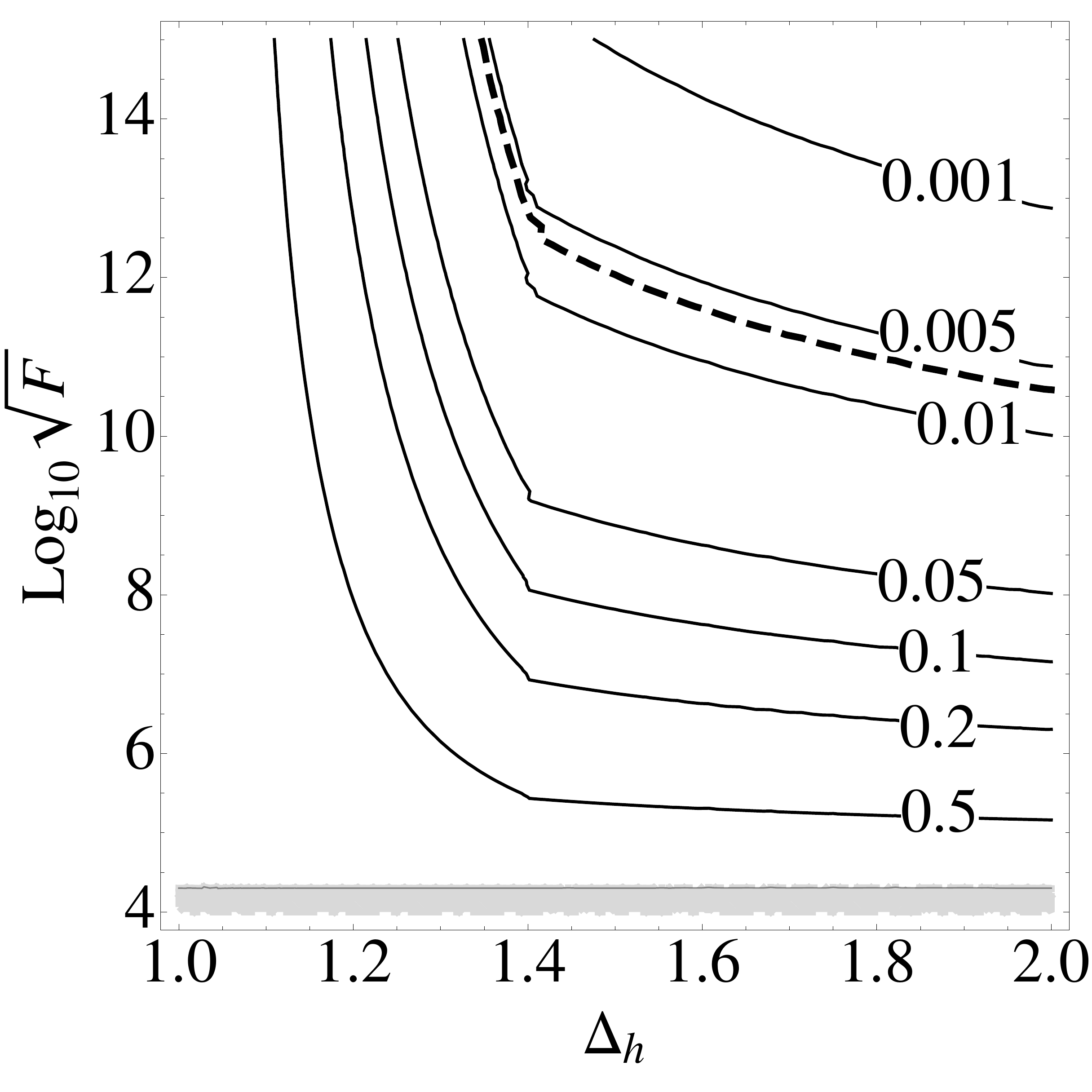}}\hfill
\subfigure[$\epsilon=0.1$]{\includegraphics[width=0.45\textwidth]{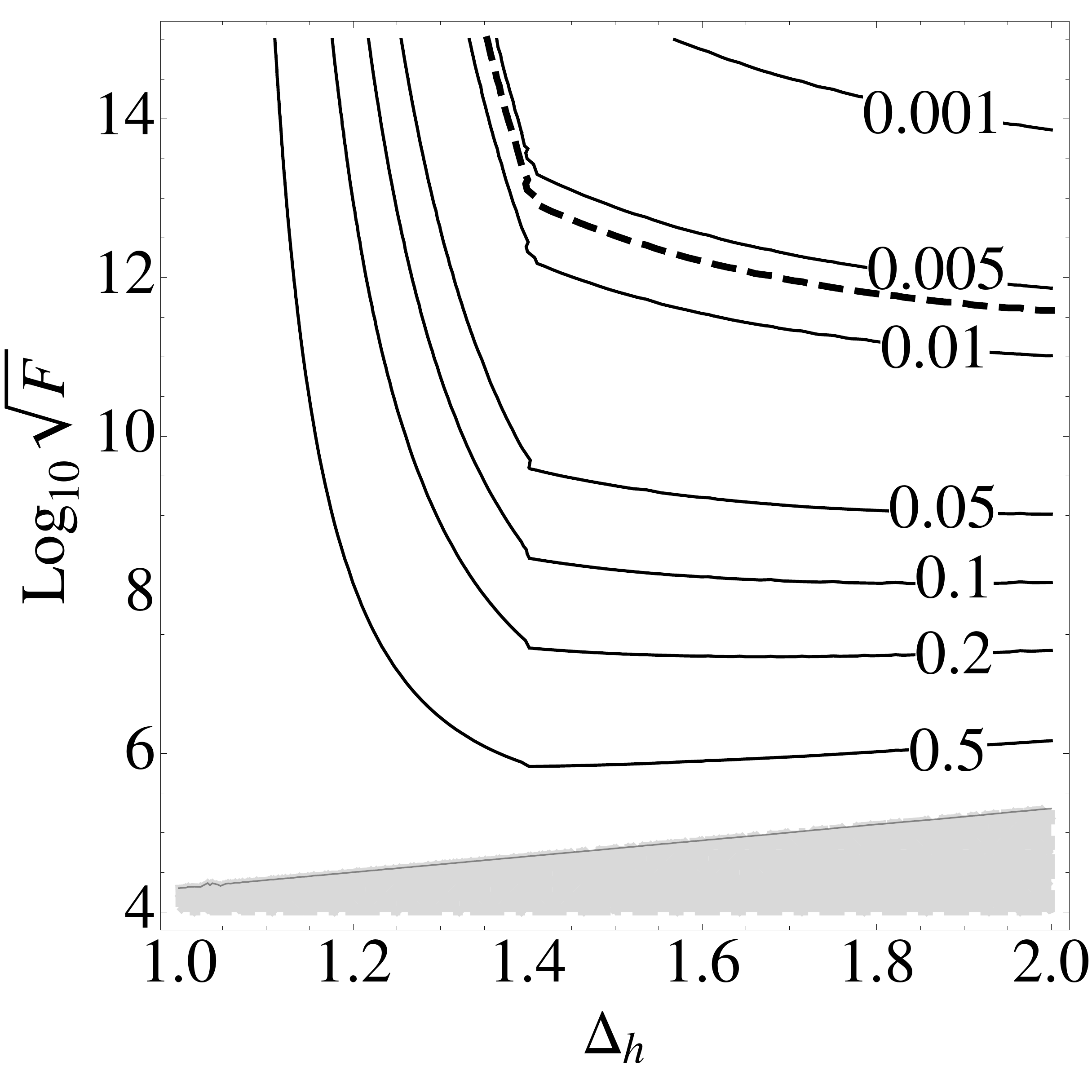}}
\caption{Contours of the maximal suppression factor that can be achieved from conformal sequestering as a function of $\Delta_h$ and $\sqrt{F}$ for various values of $\epsilon$ with $N=5$. The dashed contour indicates the suppression needed to precisely overcome the factor $16\pi^2$ that constitutes the $\mu$/$B_\mu$ and $A$/$m_H^2$ problems. The gray region corresponds to the unphysical case $\sqrt{F}>M$. The contours should be taken as a rough estimate using (\ref{gauginoschem}). The precise value of the sequestering is model-dependent.   \label{seqcontourplots}}
\end{figure}

To see this, consider some reference values in table \ref{anomalousdim}, taken from figure 7 of \cite{Poland:2011ey}. The bounds are clearly very strong for low values of $\Delta_h$, but could going to larger $\Delta_h$  allow for enough sequestering? (This is indeed suggested by figure 9 of \cite{Poland:2011ey}.) In fact, increasing $\Delta_h$ runs into a competing constraint. Because the messenger sector portal (\ref{eq:Whm}) becomes a higher-dimension operator, it becomes increasingly challenging to achieve realistic gaugino masses.\footnote{An identical argument applies to $\mu$ and $A_{u,d}$, which implies that this constraint cannot be be simply circumvented by arranging the gaugino masses to arise from a separate source of supersymmetry breaking. } 
These are given by (\ref{GGMforms}) and (\ref{Bcor}):
\begin{align}
M_i=\frac{g_i^2}{16\pi^2}N \frac{\sqrt{F}^{\Delta_h+1}}{M\Lambda^{\Delta_h-1}}.
\end{align}
Requiring TeV-scale gaugino masses leads to the following rough limit on the suppression that can be achieved from conformal sequestering:
\begin{align}
\left(\frac{\sqrt{F}}{M}\right)^\gamma\gtrsim \left(\frac{100\;\mathrm{TeV}}{N \epsilon^{\Delta_h-1}\sqrt{F}}\right)^{\frac{\gamma}{\Delta_h}}\label{gauginoschem}
\end{align}
where we made the rough order of magnitude estimate $\frac{16\pi^2}{g_i^2}M_i\sim 100$ TeV and defined $\epsilon \equiv\frac{M}{\Lambda}<1$. Figure \ref{seqcontourplots} shows this as a function of $\sqrt{F}$ and $\Delta_h$ (with $\gamma$ saturating the bootstrap bound), for two choices of $\epsilon$. For larger $\Delta_h$, the hierarchy between the messenger and hidden sector scales is greatly reduced in comparison to the hierarchy that one would obtain in the spurion limit. In combination with the upper bound on $\gamma$ from \cite{Poland:2011ey}, this severely limits the amount of sequestering that can be achieved. Some more comments on this result:
\begin{itemize}
 
\item Equation (\ref{gauginoschem}) and the requirement that $M>\sqrt{F}$ also provide a rough lower bound on $\sqrt{F} \gtrsim \frac{100\;\mathrm{TeV}}{ N\epsilon^{\Delta_h-1}}$. This is the same type of lower bound as found for any model with weakly coupled messengers, except that in the case at hand the bound is further strengthened for smaller values of $\epsilon$ and larger values of $\Delta_h$.

\item For $\Delta_h\gtrsim 1.7$, increasing $\Delta_h$ barely improves the sequestering, because of the competing effects described above. 

\item The estimate in (\ref{gauginoschem})  also shows that the achievable sequestering somewhat improves for higher $N$, however the gain is limited due to the $\sqrt{F}$-dependent upper bound on $N$ from Landau poles in the gauge couplings.

\end{itemize}

\begin{table}[t]\centering
\begin{tabular}{|c|c|}\hline
$\qquad\Delta_h \qquad $&$\qquad(\gamma)_{max} \qquad $\\\hline
1.20&0.1\\
1.45&0.4\\
2.00&0.7\\\hline
\end{tabular}
\caption{Maximum allowed value of $\gamma$ for selected values of $\Delta_h$, as extracted from figure 7 of \cite{Poland:2011ey}.  \label{anomalousdim}}
\end{table}

By comparing figure \ref{seqcontourplots} with figure 9 of \cite{Poland:2011ey}, we see that the bound on the sequestering has been strengthened considerably by accounting for TeV scale gaugino masses and by factoring in the UV scale $\Lambda$, here parametrized by the variable $\epsilon$. In particular, a full loop factor suppression is only feasible for $\sqrt{F}\gtrsim10^{11}$ GeV. In this case the separation with the weak scale may be sufficiently high such that MSSM RG-running could suffice to generate a large $A$-term, without the need for Higgs mediation \cite{Draper:2011aa}. Moreover, such a high scale of supersymmetry breaking introduces various subtleties in the model: firstly, the gravitino is no longer the LSP. While this is interesting if the new LSP is a neutralino \cite{Murayama:2007ge,Craig:2008vs,Craig:2009tz}, it is a disaster if the new LSP is a stau.  Secondly, for $\sqrt{F}\gtrsim 10^{11}$ GeV, contributions from anomaly and/or gravity mediation may not be negligible. Especially the latter could be problematic, as they generically induce large flavor violation in the $A$-terms and the sfermion masses. (On the other hand, it is possible that the very same mechanism of conformal sequestering may help to suppress dangerous flavor violation \cite{Randall:1998uk,Luty:2001jh,Luty:2001zv}.) While these are certainly interesting issues, we do not wish to confront them in this paper. For simplicity we therefore restrict our discussion to $\sqrt{F}<10^{10}$ GeV, to ensure that the gravitino is always the NLSP and that gravity-induced flavor violation is always automatically small. 

From figure \ref{seqcontourplots} we then conclude that for \mbox{$\sqrt{F}<10^{10}$ GeV} conformal sequestering is not sufficiently powerful to achieve the fully suppressed boundary conditions in (\ref{fullyseq}). The best we can hope for is to achieve a partial suppression from  sequestering given by 
\begin{align}
\left(\frac{\sqrt{F}}{M}\right)^\gamma\sim 0.01-0.1.
\end{align}
This may be sufficient -- especially in combination with some additional suppression from $\hat C<1$ and/or $N>1$ -- to achieve viable EWSB, provided that the boundary conditions at the scale $\sqrt{F}$ still satisfy
\begin{align}
B_\mu\lesssim |\mu|^2 \quad\mathrm{and}\quad \hat m_{H_u}^2 \lesssim |A_u|^2,
\end{align}
rather than the overly stringent requirement in (\ref{fullyseq}). Such partially suppressed boundary conditions imply that the details of the dynamics in the hidden and messenger sectors are not erased at the scale $\sqrt{F}$. Instead, both sectors should leave an observable imprint on the low energy spectrum. Using the GMHM formalism developed in \cite{Craig:2013wga}, we are able for the first time to explicitly evaluate this imprint for a weakly messenger sector of our choice. We will present an explicit example in section \ref{example}, but before doing so, it is useful to study the available parameter space in a (semi) model-independent way. This will be the subject of the next section.

\FloatBarrier

\section{Exploring the Parameter Space\label{secparamspace}}
The correlator formalism described in the previous section a priori involves a very large parameter space; in the most general case the boundary conditions are described by no less than 10 free parameters:
\begin{equation}\label{paramspace}
M_{1},\,\, M_{2},\,\, M_{3},\,\,  A_u,\,\, A_d,\,\, \mu,\,\, B_\mu,\,\, \mhu^2,\,\, \mhd^2,\,\, \sqrt{F}.
\end{equation}
with $m^2_{\tilde f}\approx 0$. (Recall from the discussion around equation (\ref{eq:gauginosfermion}) that the sfermion masses are suppressed at the scale $\sqrt{F}$.) Here and onwards, the parameters in (\ref{paramspace}) are always to be thought of as evaluated at the scale $\sqrt{F}$, unless indicated otherwise. Following the discussion of the previous section, to maximize the impact of the conformal sequestering we choose $\sqrt{F}=10^9$ GeV.  At the end of section \ref{example} we briefly comment on lower values for $\sqrt{F}$.

\begin{figure}[t!]\centering
\includegraphics[width=1 \textwidth]{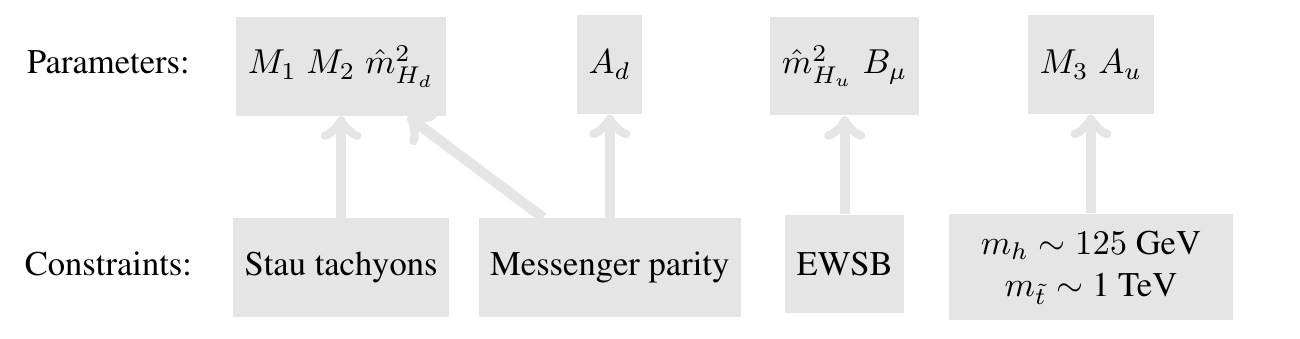}
\caption{Schematic representation of the various constraints and how they impact the parameter space. We use the electroweak symmetry breaking conditions to eliminate $B_\mu$ in favor of $\tan\beta$. Our assumption regarding the action of the messenger parity on the operators $O_u$ and $O_d$ allows us to eliminate $A_d$ as an independent variable and to constrain $\mhd$ to be positive. \label{flowchart}}
\end{figure}

Before even writing down a specific UV model, we can restrict this parameter space through  phenomenological considerations in the IR such as EWSB and the Higgs mass. This approach has a double advantage: it serves as a valuable intermediate step in the full analysis and provides some model independent information about the UV soft parameters. Despite the restrictions from the IR boundary conditions, the remaining parameter space in (\ref{paramspace}) is still rather daunting to analyze in full generality. In this paper, we instead choose to impose one more condition on the UV soft parameters purely for simplicity. This condition -- an extension of messenger parity to the Higgs-messenger portal -- renders the parameter space in (\ref{paramspace}) manageable. Moreover it is a property of a broad class of models, and it is motivated in particular by the model we will study in section \ref{example}. The impact of each of the constraints on the soft parameters is summarized in figure \ref{flowchart}, and in this section we will describe each one in turn.

\subsection{Simplifying assumptions for the UV soft parameters}
\label{subsec:UVbdyconditions}

In GGM, a standard ingredient is that the hidden sector possesses a ``messenger parity" symmetry that forbids dangerous hypercharge tadpoles \cite{Dimopoulos:1996ig,Meade:2008wd}. To reduce the size of the parameter space here, we choose to 
extend this symmetry to the Higgs-messenger interactions. Specifically, we assume that messenger parity exchanges $O_u$ and $O_d$. This greatly simplifies our analysis, since it implies that the correlators for $A_u$ and $A_d$ in (\ref{eq:A}) must be identical. The same is true for the correlators for $\mhu^2$ and $\mhd^2$ in (\ref{mhcor}). The soft parameters must therefore obey the following relation at the scale $\sqrt{F}$:
\begin{equation}
\frac{A_d}{A_u}=\frac{\mhd^2}{\mhu^2}=\frac{|\lambda_{d}|^2}{|\lambda_{u}|^2}>0.\label{Adrelation}
\end{equation}
We can conveniently use this constraint to eliminate $A_d$ as a free parameter, and thus reduce size of the parameter space. In addition, (\ref{Adrelation}) determines the relative sign of $A_u$ and $A_d$, as well as the relative sign of $\mhu^2$ and $\mhd^2$. We emphasize that this extension of messenger parity to $O_u$ and $O_d$ is motivated purely on the grounds of convenience; although  messenger parity is usually included in the definition of gauge mediation, in general it does not need to act on $O_u$ and $O_d$ in this specific way.

For any concrete model, the UV soft parameters must be realized in terms of the underlying parameters of the model, which generally leads to additional restrictions on top of (\ref{Adrelation}). For example, a minimal messenger sector with only messengers in a ${\bf 5}$-${\bf \bar 5}$ representation of an SU(5) GUT yields the following relation between the gaugino masses:
\begin{equation}
M_1=\frac{3}{5}\frac{g_1^2}{g_2^2}M_2+\frac{2}{5}\frac{g_1^2}{g_3^2}M_3. \label{gauginorelation}
\end{equation}
In this section we discuss this special case as well as the more general case where all three gaugino masses are independent. Any further restrictions on the UV boundary conditions are typically highly model-dependent, and we deal with them only when we commit to a specific example in section \ref{example}. 

\subsection{IR boundary conditions}

The restrictions on the IR soft masses are purely given by phenomenological considerations, and as such they are independent of the precise composition of the messenger sector. In particular, we demand that a realistic spectrum at the weak scale satisfies the following requirements:
\begin{enumerate}
\item Viable electroweak symmetry breaking.
\item $m_h\approx 126$ GeV and TeV-scale stops.
\item Charge, color and CP must be unbroken in the vacuum on cosmological time scales.
\end{enumerate} 
In what follows, we will go step by step through the IR constraints mentioned above, and use them to reduce the size of the parameter space until it becomes tractable. More details on our numerical procedure are given in appendix \ref{Appnumerical}.

\subsubsection{Constraints from EWSB}

As usual, the tadpole equations in the Higgs sector allow us to eliminate $\mhu^2$ and $B_\mu$ at the weak scale in favor of $m_Z$ and $\tan\beta$. In order not to exacerbate the fine-tuning, we only consider $|\mu|\leq500$ GeV, but this is by no means essential. This assumption has the additional benefit that the parameter $\mu$ now has little impact on the IR spectrum, with the exception of course of the mass of the Higgsino, which may be the NLSP. A small number of discrete choices therefore suffices to obtain a good qualitative picture of the parameter space. In addition, we fix $\tan\beta=10$. 

\subsubsection{Constraints from the Higgs mass\label{sec:higgsmass}}

\begin{figure}[p]\centering
\subfigure[$M_2=0.5$ TeV]{\includegraphics[height=7cm]{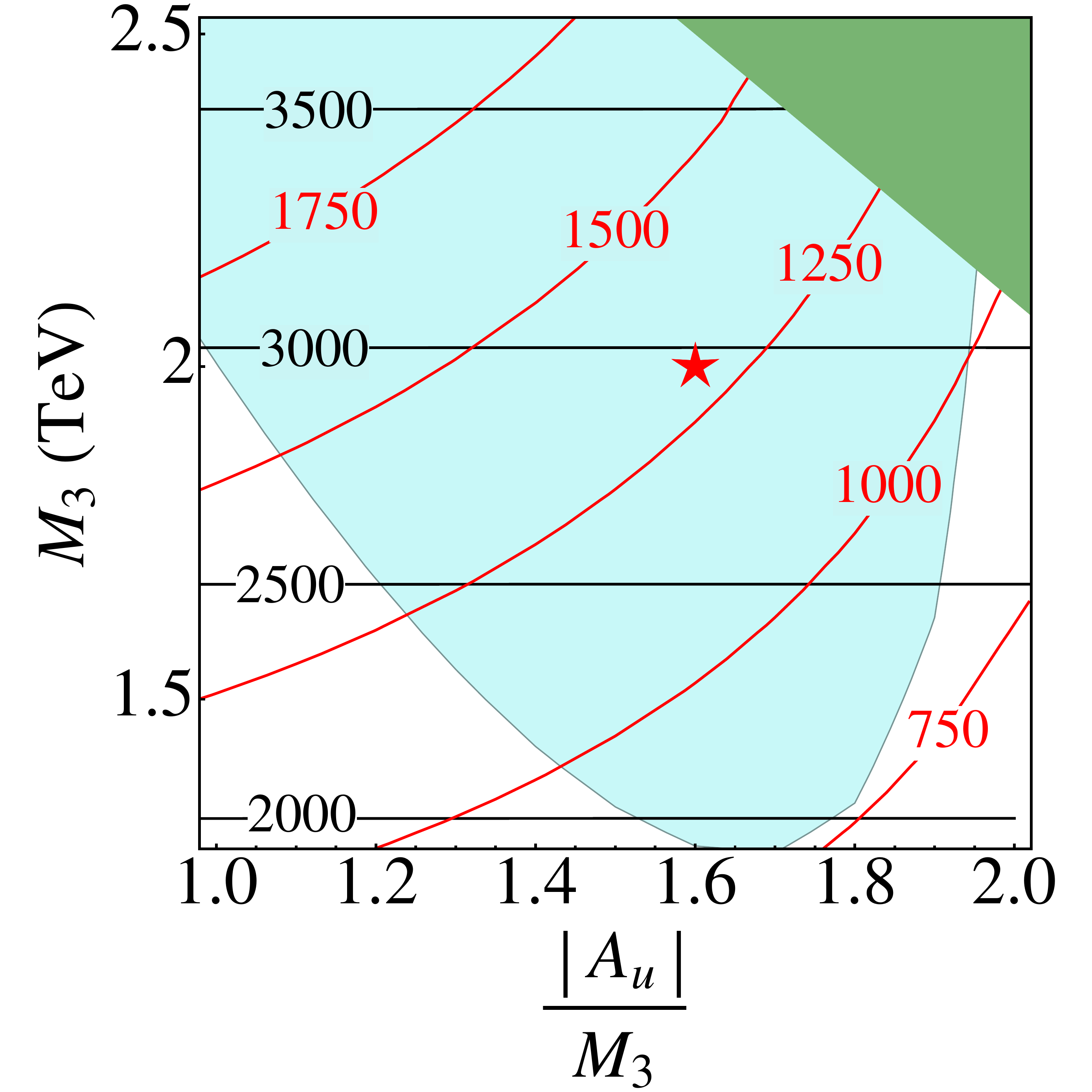}}
\subfigure[$M_2=2$ TeV]{\includegraphics[height =7cm]{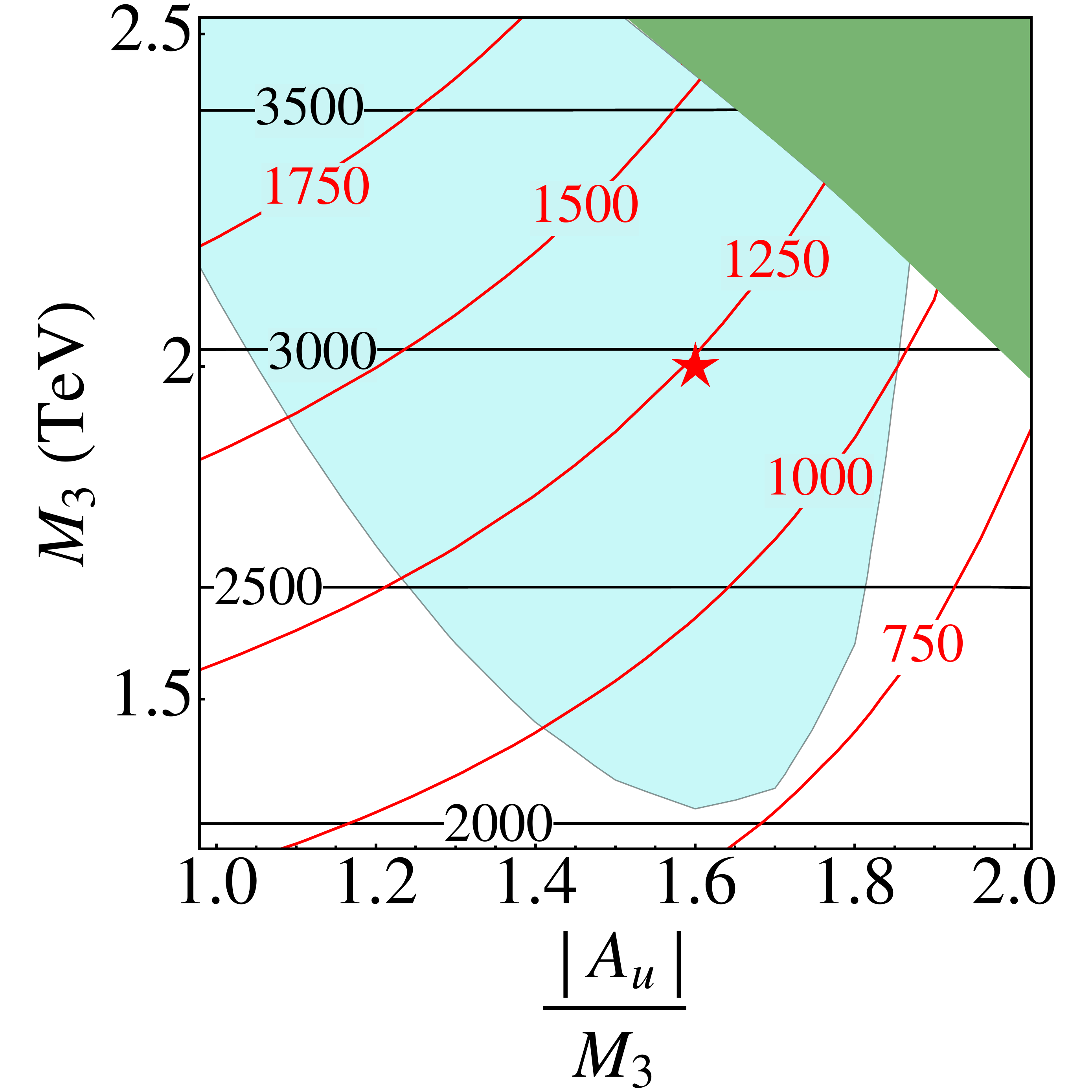}}
\subfigure[${\bf 5}$-${\bf \bar 5}$ messengers]{\includegraphics[height =8cm]{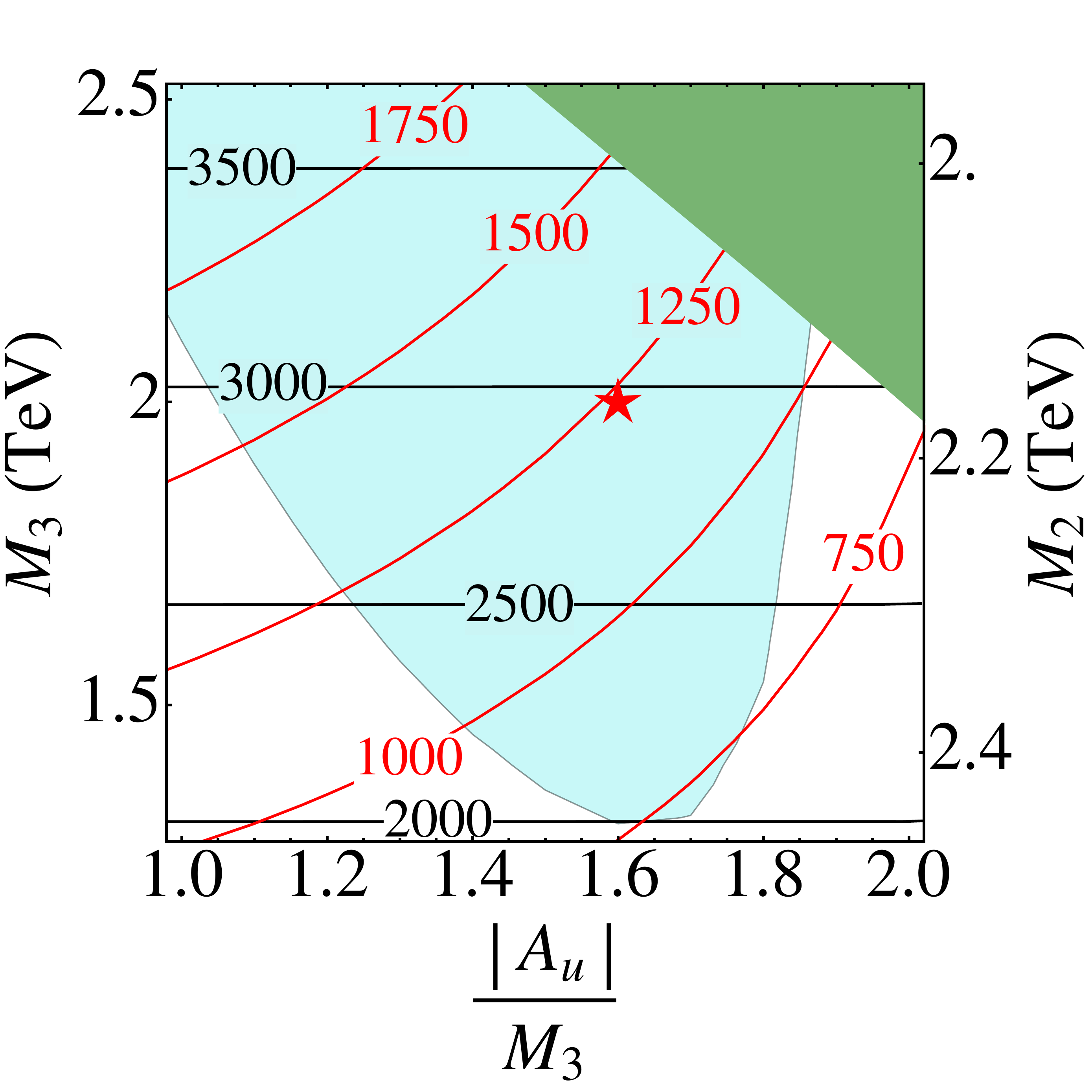}}
\caption{Contours of the pole masses (in GeV) of the lightest stop (red) and the gluino (black), as a function of $M_3$ and $|A_u|/M_3$, for different choices of $M_2$. The other parameters are fixed to $\tan\beta=10$, $\mu=200$ GeV, $\sqrt{F}=10^9$ GeV, $M_1=1.2$ TeV and $m_{A^0}=1.5$ TeV. The pseudoscalar pole mass $m_{A^0}$ was used instead of $\hat m_{H_d}^2$ for purely technical reasons; all other parameters are defined at the scale $\sqrt{F}$. The blue region represents $123$ GeV $<m_h<$ 129 GeV; the green region is ruled out by stau tachyons. The star indicates the benchmark point plotted in figure \ref{general_master_plot}.\label{moneyplot1} }
\end{figure}

The parameters $A_u$ and $M_3$ are the most important parameters as far as the mass of the lightest CP even Higgs is concerned, as they set the stop $A$-term as well as the stop masses. To appreciate the latter, consider the system of RG equations
\begin{align}
16\pi^2\frac{d}{dt}m^2_{Q_3}&=2y_t^2(\mhu^2+m^2_{Q_3}+m^2_{u_3}+|A_u|^2)-\frac{32}{3}g_3^2|M_3|^2-6g_2^2|M_2|^2\nonumber\\
16\pi^2\frac{d}{dt}m^2_{u_3}&=4y_t^2(\mhu^2+m^2_{Q_3}+m^2_{u_3}+|A_u|^2)-\frac{32}{3}g_3^2|M_3|^2\label{stoprge}
\\
16\pi^2\frac{d}{dt}\mhu^2&=6y_t^2(\mhu^2+m^2_{Q_3}+m^2_{u_3}+|A_u|^2)-6g_2^2|M_2|^2\nonumber
\end{align}
where we neglected contributions proportional to $y_b$ and $g_1$. We also dropped the dependence on the $\mu$ parameter, since we assumed it to be smaller than the other soft masses. The key fact is that the stop masses and $\mhu^2$ are essentially zero at the scale $\sqrt{F}$ (due to sequestering) and at the weak scale (due to EWSB), respectively. Therefore, the running of the stops and $\mhu^2$ must be determined primarily by the sources $M_3$, $M_2$ and $A_u$. Of these parameters, the effect of $M_2$ is typically subleading compared to the other two.  It is therefore justified to fix the parameters $A_u$ and $M_3$ by insisting on $m_h\approx126$ GeV with TeV-scale stop masses.\footnote{A priori, a large $A_u$ may cause our vacuum to decay to a lower, color-breaking vacuum on a time scale shorter than the age of the universe. The recently improved empirical constraint on this process \cite{Blinov:2013fta} does not impact the parameter space plotted in figure \ref{moneyplot1}. (See also \cite{Camargo-Molina:2013pka,Chowdhury:2013dka} for similar recent results.)} This is illustrated in figure \ref{moneyplot1} for some representative values of $M_2$. Given both the theory and the experimental errors on the Higgs mass, this is necessarily a somewhat loose constraint, and for the purpose of our analysis, we simply choose a representative point in the allowed region, indicated with a star in figure \ref{moneyplot1}. Other choices are certainly possible, but the qualitative features of what will follow are preserved. 

\subsubsection{Constraints from tachyons and (meta)stability}

\begin{figure}[p]\centering

\subfigure[$M_2=0.5$ TeV\label{masterlowM2}]{\includegraphics[height=0.29\textheight]{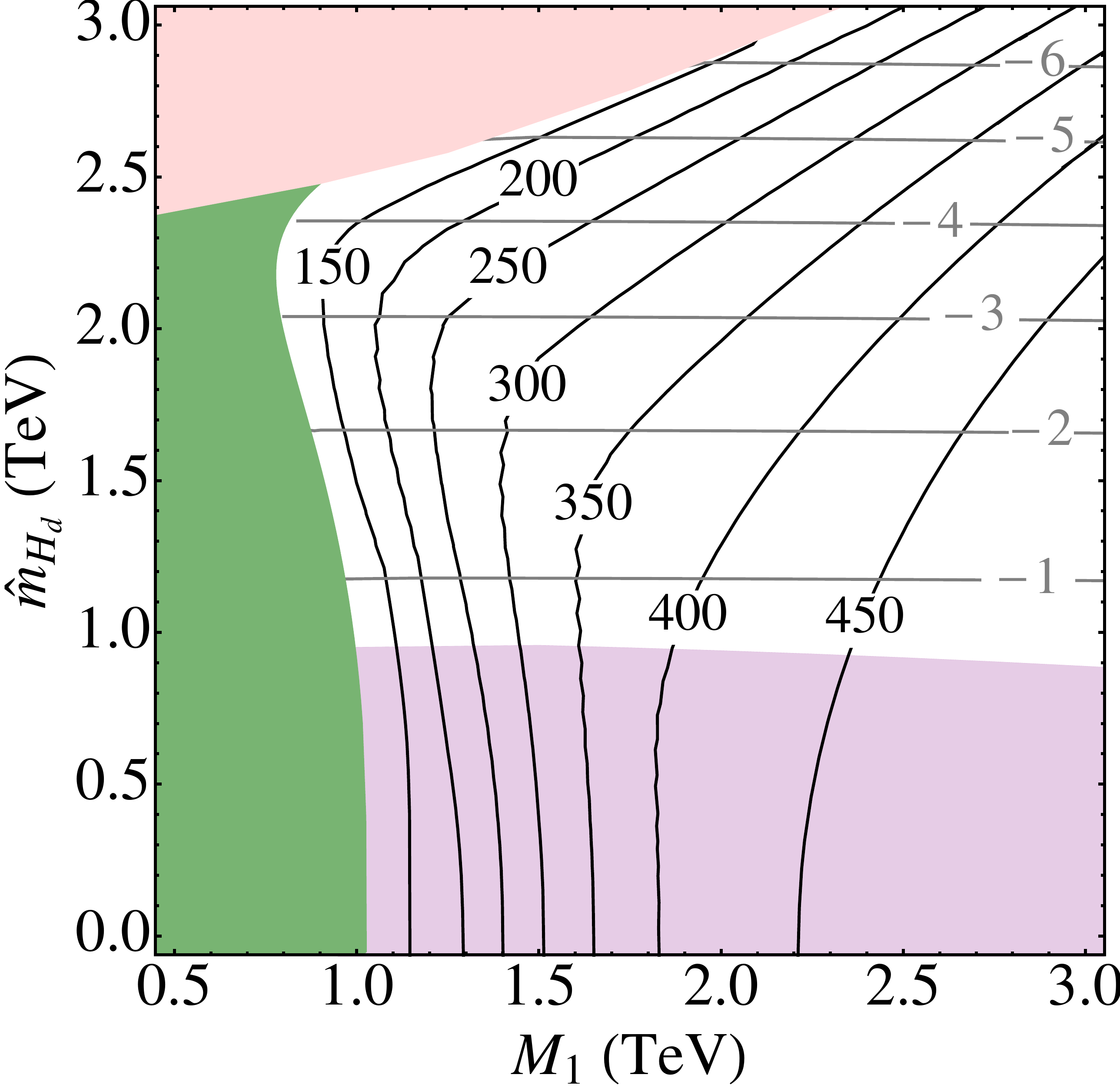}}\hfil
\subfigure[$M_2=2$ TeV\label{masterhighM2}]{\includegraphics[height =0.29\textheight]{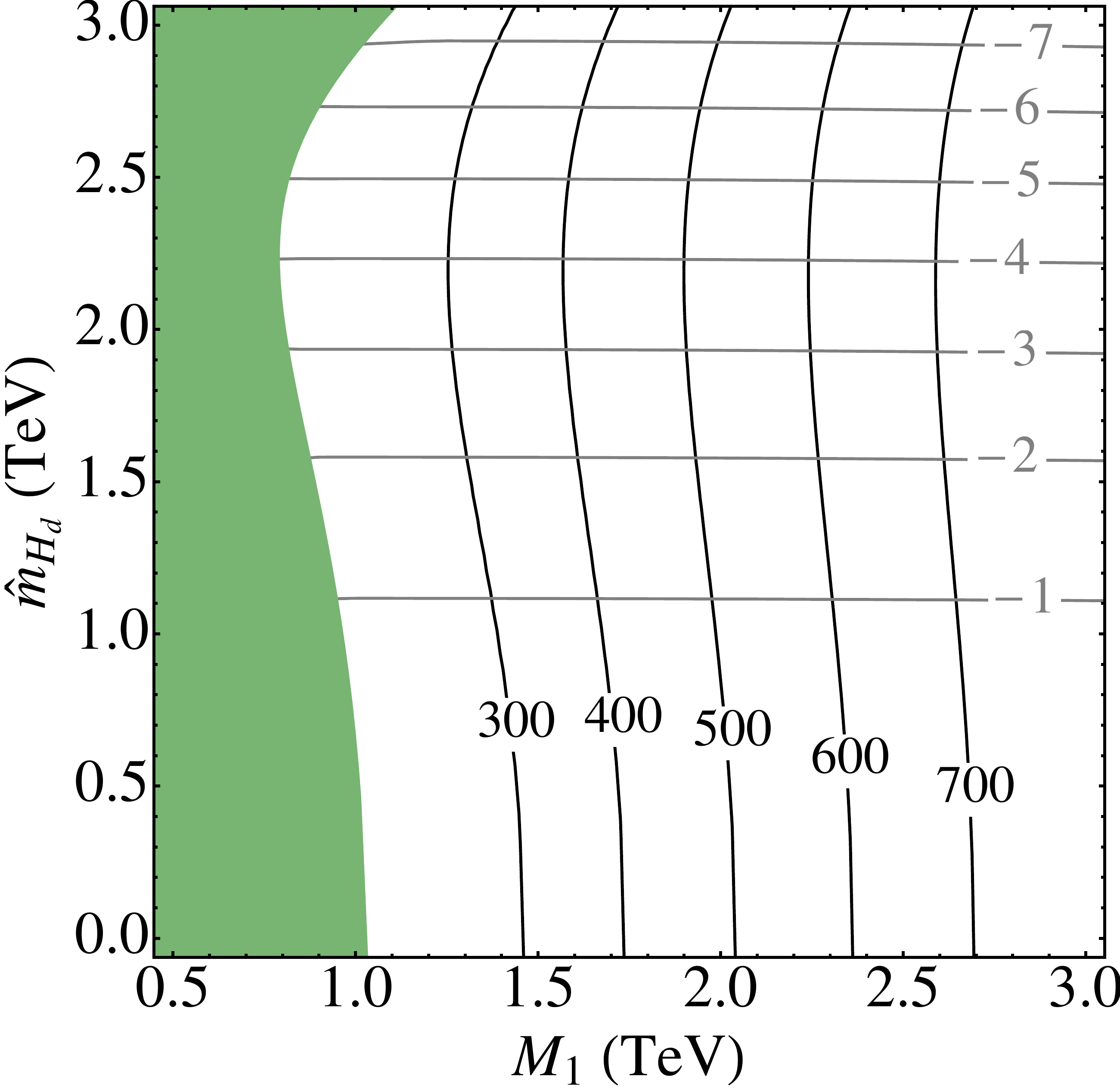}}
\subfigure[${\bf 5}$-${\bf \bar 5}$ messengers\label{mastergauginounification}]{\includegraphics[height=0.335\textheight]{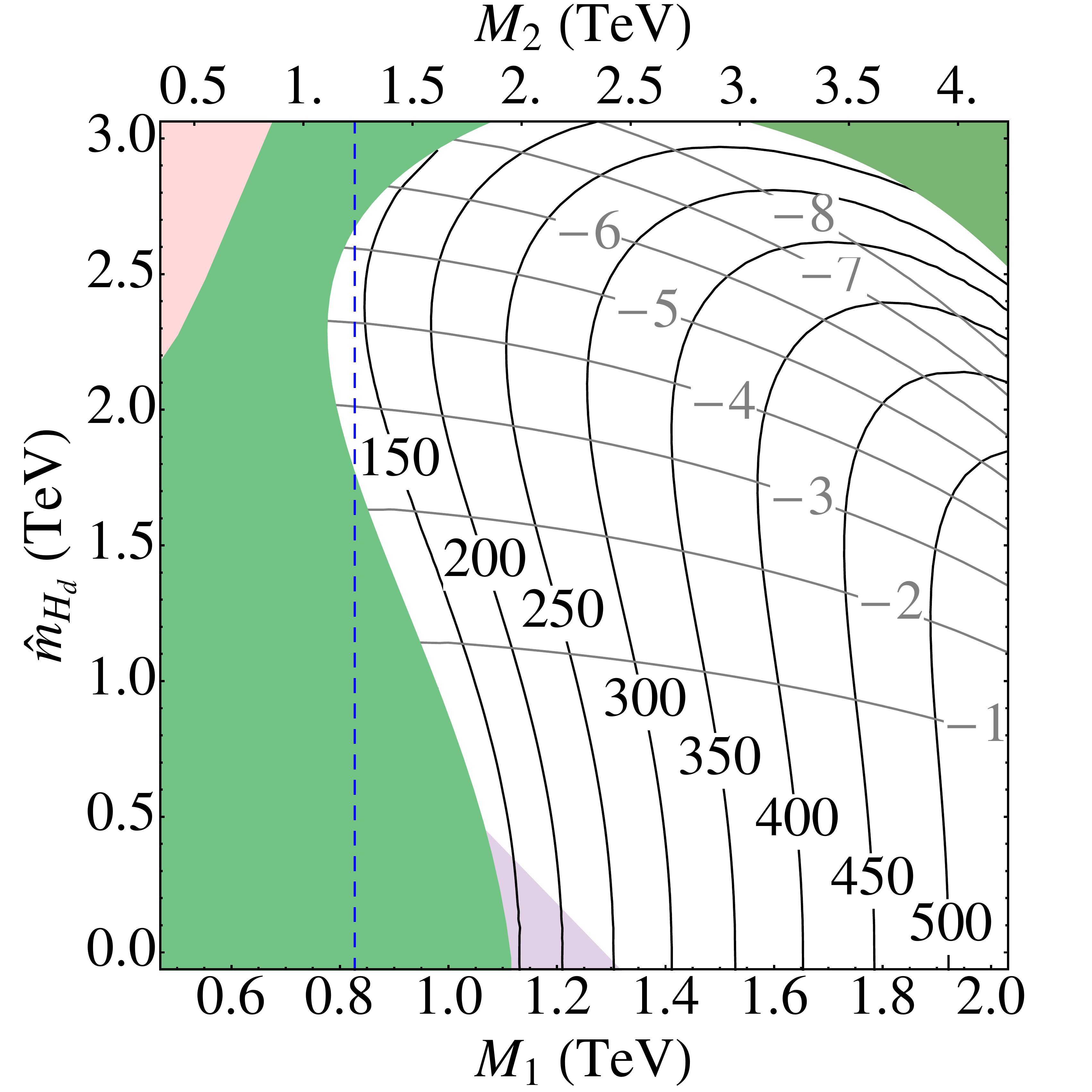}}

\caption{The pole mass of the lightest stau in GeV (black) and $A_d$ in TeV (gray) as a function of  $M_1$ and $\mhd$. $M_3=2.0$ TeV and $A_u=-3.2$ TeV and were chosen such that a $m_h\approx126$ GeV is achieved with TeV-scale stop masses (see the red star on figure \ref{moneyplot1}). The other parameters were fixed to $|\mu|=400$ GeV, $\sqrt{F}=10^9$ GeV and $\tan\beta=10$. All soft parameters are defined at the scale $\sqrt{F}$.   The green (red) shaded region indicates a stau (snutau) tachyon. If the $\mu<0$, the purple region is ruled out by an $A^0$ tachyon. The blue dashed line in \ref{mastergauginounification} indicates the slice of parameter space where the gaugino masses unify at the GUT scale.  \label{general_master_plot}}
\end{figure}

Having fixed $A_u$ and $M_3$ from requiring TeV-scale stops and $m_h\approx 126$ GeV, we are left with just the independent parameters $M_1$, $M_2$ and $\mhd^2$ (see figure \ref{flowchart}). All of these will be constrained by requiring the absence of slepton tachyons. Since the Yukawa interaction pushes the sleptons down in the RG running, the third generation is always the most constraining. 
The relevant RG equations are\footnote{Keep in mind that $\mhd^2$ does not exhibit strong RG running and can usually be approximated fairly well by its UV value. The story is very different for the stau masses: although in absolute terms their RG running is small as well, their UV threshold value is highly suppressed and the running therefore provides the dominant contribution to the IR stau masses.}  
\begin{align}
16\pi^2\frac{d}{dt}m^2_{L_3}&=2y_\tau^2 |A_d|^2-6g_2^2 |M_2|^2-\frac{6}{5}g_1^2 |M_1|^2-\frac{3}{5}g_1^2 S\label{stauL}\\
16\pi^2\frac{d}{dt}m^2_{e_3}&=4y_\tau^2 |A_d|^2-\frac{24}{5}g_1^2 |M_1|^2+\frac{6}{5}g_1^2 S\label{stauR}
\end{align}
with
\begin{align}
S&=\mathrm{Tr}[Y_i m^2_{\phi_i}].\label{FIterm}
\end{align}
At the scale $\sqrt{F}$, $S\approx \mhu^2-\mhd^2$, since all sfermion masses are small.  Given that $y_\tau^2\ll g_1^2$, we neglect all the terms proportional to $y_\tau^2$, except for $|A_d|^2$, which may be very large. The right-handed stau is the more fragile of the two staus, since its mass is not sensitive to the upwards pull of $M_2$. Moreover recall that $\mhu^2$ and $A_d$ have already been fixed by EWSB plus the Higgs mass constraint and the extension of messenger parity, respectively. The most interesting slicing of the parameter space is therefore in terms of $M_1$ and $\mhd^2$. This is shown in the plots in figure \ref{general_master_plot}, and we now proceed to discuss these plots in more detail.\footnote{Note that in these plots we have considered only positive $\mhd^2$. This is because  $\mhu^2$ (at the scale $\sqrt{F}$) is positive for our choices of $M_3$, $M_2$ and $A_u$, and our simplifying assumption about messenger parity relates the sign of $\mhd^2$ to that of $\mhu^2$ through equation (\ref{Adrelation}). }

Let us first consider the case where $M_2$ is held fixed, as shown in figure \ref{masterlowM2} and figure \ref{masterhighM2} for two representative values of $M_2$.  

\begin{itemize} 

\item Since $\mhd^2$ pushes down $m_{L_3}^2$ in the RGE, as $\mhd^2$ is increased, it eventually results in a snutau tachyon. This is indicated by the red shaded region in figure \ref{masterlowM2}. This is less of an issue for larger $M_2$, which is why there is no analogous constraint from snutau tachyons in figure \ref{masterhighM2}.

\item For smaller $M_1$, either the $S$-term or the $|A_d|^2$ term drives the right-handed stau tachyonic. This is indicated by the green shaded regions in figure \ref{general_master_plot}.

\item Another interesting feature in figure \ref{masterlowM2} and figure \ref{masterhighM2} is that $A_d$ is fairly independent of $M_1$ and monotonically increases as a function of $\mhd^2$. This is a direct consequence of (\ref{Adrelation}) and the fact that (as we just discussed) $\mhu^2$ is basically constant in these plots.

\item A final noteworthy special case occurs if $M_2\ll M_1$, as the lightest stau mass eigenstate may be predominantly composed out of the left-handed stau, due to the smaller coefficient for the $|M_1|$ term in (\ref{stauL}) compared to its analogue in (\ref{stauR}). 

\end{itemize}

In models with only ${\bf 5}$-${\bf \bar 5}$ messengers, $M_2$ is a function of $M_1$ and $M_3$ rather than an independent variable. The constraints on this case are shown in figure \ref{mastergauginounification}.

\begin{itemize}

\item We see from figure \ref{mastergauginounification} that $M_2$ is always larger than $M_1$, so snutau tachyons no longer constrain the parameter space, as a stau tachyon is always generated first. 

\item The relation between the gaugino masses has some interesting implications on $A_d$ and the lightest stau as shown in figure \ref{mastergauginounification}. In particular, the $A_d$ contours bend downwards for large values of $M_2$. This is again easily understood from (\ref{Adrelation}) and (\ref{stoprge}): for large values of $M_2$, $\mhu$ is smaller at the scale $\sqrt{F}$, which in turn leads to a large and negative $A_d$. Since $A_d$ pulls the staus down, the stau contours eventually start tracking the $A_d$ contours for sufficiently large $A_d$, and ultimately a stau tachyon is induced. Interestingly, this leads to an {\it upper} bound on $M_1$ from stau tachyons, a priori a somewhat counterintuitive notion. 

\item Also note that the special scenario where all the gaugino masses unify at the GUT scale (dashed blue line in figure \ref{mastergauginounification}) is only viable in a small sliver of the parameter space for $\mhd\sim 2$ TeV. 

\end{itemize}

Finally, we verified using \texttt{Vevacious-1.0.11}  \cite{Camargo-Molina:2013qva}
that there are no further significant constraints from metastable vacuum decay to a charge breaking minimum (even with such large $A_d$).  However, the parameter space is constrained by demanding the absence of CP-breaking vacua. If $\mu<0$ and $|A_u|\gg M_2$, the pseudoscalar may end up tachyonic by virtue of a large radiative correction. This constraint is indicated by the purple region in figure \ref{general_master_plot}.

\subsection{Summary of the constraints}

This concludes the discussion of the (semi) model-independent constraints on the parameter space. Since the discussion was rather lengthy and involved, a brief summary is appropriate:

\begin{itemize}
\item Our assumptions on the extension of messenger parity let us eliminate $A_d$ as a free parameter through (\ref{Adrelation}) and restrict $\mhd^2$ to be positive at the scale $\sqrt{F}$.
\item Through the EWSB conditions we eliminate $\mhu^2$ and trade $B_\mu$ for $\tan\beta$. Except for the Higgsino mass, the IR physics has little sensitivity to the $\mu$ parameter.
\item Requiring $m_h\approx 126$ GeV for a minimal SUSY scale roughly fixes $M_3$ and $A_u$. As an extra consequence, this requirement also more or less determines $\mhu^2$ at the scale $\sqrt{F}$.
\item The absence of a charge and CP breaking vacuum imposes restrictions on the parameters $M_1$, $M_2$ and $\mhd^2$. Roughly speaking, this leads to a lower bound on $M_1$ and an upper bound on $\mhd^2$.
\end{itemize}

Now that we have exhausted all (semi) model-independent constraints, we will write down an explicit example and compute the associated UV boundary conditions using GMHM. These boundary conditions then yield a prediction for the conformal sequestering and the effective OPE coefficient $\hat C$.

\FloatBarrier

\section{A Minimal Example\label{example}}

Perhaps the simplest example of a messenger sector which generates both $\mu$ and $A_u$ at one loop is  
\begin{align}
W=\left(\frac{\kappa}{\Lambda^{\Delta_h-1}}O_h +M\right)\left(\tilde\phi_D\phi_D+ \tilde\phi_S\phi_S\right)+\lambda_{u}\tilde\phi_D\phi_S H_u+\lambda_{d} \phi_D\tilde\phi_S H_d.\label{Wproxy}
\end{align}
where the $\phi_D$ and $\phi_S$ are a $SU(2)$ doublet and a gauge singlet respectively. In the spurion limit this model notoriously yields the disastrous relation $B_\mu\sim 16\pi^2 \mu^2$ \cite{Dvali:1996cu}. However, as we will show, when hidden sector effects are accounted for this is not necessarily the case. 

To obtain a complete model we embed the doublet messengers in ${\bf 5}$-${\bf \bar 5}$ representations of $SU(5)$ and exploit the full parametric freedom of the model. The full superpotential is then
\begin{align}
W=&\frac{O_h}{\Lambda^{\Delta_h-1}}\left(\kappa_{T}\tilde\phi_T\phi_T+\kappa_{D}\tilde\phi_D\phi_D+\kappa_{S}\tilde\phi_S\phi_S\right)+M_{T}\tilde\phi_T\phi_T+M_{D}\tilde\phi_D\phi_D+M_{S}\tilde\phi_S\phi_S\nonumber\\
&+\lambda_{u}\tilde\phi_D\phi_S H_u+\lambda_{d} \phi_D\tilde\phi_S H_d\label{Wsimple}
\end{align}
where the $\phi_T$, $\tilde\phi_T$ are $SU(3)$ triplets. Note that they do not participate in the Higgs mediation; their sole purpose is to complete the $SU(5)$ multiplet and to give a mass to the gluino through standard gauge mediation.  $M_T$, $M_D$ and $M_S$ can all be chosen positive without loss of generality. As is conventional, we allow for $N$ identical copies of these messengers, as long as no Landau poles are introduced below the GUT scale. 

\subsection{UV boundary conditions}
The threshold contributions to the gaugino masses are the usual ones in gauge mediation, and may be obtained from (\ref{GGMforms}) and (\ref{Bcor}):
\begin{align}
M_3&=\frac{g_3^2}{16\pi^2}  \Lambda_T\nonumber\\
M_2&=\frac{g_2^2}{16\pi^2}    \Lambda_D \label{bino} \\
M_1&=\frac{3}{5}\frac{g_1^2}{16\pi^2}    \Lambda_D+\frac{2}{5}\frac{g_1^2}{16\pi^2}   \Lambda_T\nonumber
\end{align}
with 
\begin{equation}
\Lambda_{D,T}=N \kappa_{D,T}\frac{\sqrt{F}^{\Delta_h+1}}{M_{D,T}\Lambda^{\Delta_h-1}}.\label{Ldtdefinition}
\end{equation}
Since the messenger sector consists out of ${\bf 5}$-${\bf \bar 5}$ messengers, the only two out of the three gaugino masses are independent and the relation in (\ref{gauginorelation}) is satisfied. 

At one loop, the threshold corrections to the Higgs sector obtained from integrating out (\ref{Wsimple}) are symmetric under interchange of $(\kappa_S,\,M_S) \leftrightarrow (\kappa_D,\,M_D)$. This symmetry is made manifest if we introduce the notation:
\begin{align} 
\kappa=\sqrt{\kappa_D\kappa_S},\quad M=\sqrt{M_D M_S} ,\quad a=\sqrt{\frac{M_D}{M_{S}}},\quad b =\sqrt{\frac{\kappa_D}{\kappa_S}}
 \end{align}
 and
 \begin{align}
 \Lambda_H\equiv N\kappa \frac{\sqrt{F}^{\Delta_h+1}}{M\Lambda^{\Delta_h-1}}
 \end{align}
Then the symmetry becomes $a\to 1/a$, $b\to 1/b$ with $\kappa$, $M$ and $\Lambda_H$ unchanged. The soft parameters can be written as:
\begin{align}
\mu&=\frac{\lambda_u\lambda_d}{16\pi^2}  f_\mu(a,b)\Lambda_H\label{simplemu}\\
A_{u,d}&=\frac{|\lambda_{u,d}|^2}{16\pi^2}  f_A(a,b)  \Lambda_H\label{simpleA}
\end{align}
The dimensionless functions $f_\mu$ and $f_A$ can be obtained from explicit computation of the appropriate correlation functions in section \ref{secreview}:
\begin{align}
f_\mu(a,b)&=\frac{a b}{\left(a^4-1\right)^2}\big(1-a^4+4  \log a\big)+(a\leftrightarrow \frac{1}{a},b\leftrightarrow \frac{1}{b})\\
f_{A}(a,b)&=\frac{a^3 b}{\left(a^4-1\right)^2}\big(1-a^4+4  \log a\big)+(a\leftrightarrow \frac{1}{a},b\leftrightarrow \frac{1}{b})
\end{align}
Similarly, the dimension two soft parameters are given by
\begin{align}
B_\mu&=\frac{\lambda_u\lambda_d}{16\pi^2} f_{B}(a,b,\gamma)\;  \frac{\hat C}{N}  \left(\frac{\sqrt{F}}{M}\right)^\gamma\Lambda_H^2\label{simpleBmu}\\
\mhud^2&=\frac{|\lambda_{u,d}|^2}{16\pi^2}   f_{m_H}(a,b,\gamma)\;  \frac{\hat C}{N} \left(\frac{\sqrt{F}}{M}\right)^\gamma\Lambda_H^2\label{simplemH}
\end{align}
where $\hat C$ is the effective OPE coefficient as defined in section \ref{secreview}, the suppression factor $\left(\frac{\sqrt{F}}{M}\right)^\gamma$ is the result of the conformal sequestering, and
\begin{align}
f_{B}(a,b,\gamma)=&\frac{\pi ^{3/2} \csc \left(\frac{\pi  \gamma }{2}\right) \Gamma \left(\frac{\gamma }{2}+2\right) \Gamma \left(\frac{\gamma }{2}\right)}{4 \left(a^4-1\right)^2 \Gamma \left(-\frac{\gamma }{2}\right) \Gamma \left(\frac{\gamma +3}{2}\right)} \Bigg(b^2 a^{-\gamma }  \left(\gamma -a^4 (\gamma +4)\right)+2 \left(a^4+1\right) a^{ \gamma +2}\nonumber\\&
-2 \left(a^4-2 a^2 b^2+1\right) a^{ \gamma +2} \, _2F_1\left(\frac{\gamma }{2},\frac{\gamma +2}{2};\gamma +2;1-a^4\right)\Bigg)+(a\leftrightarrow \frac{1}{a},b\leftrightarrow \frac{1}{b})\\
f_{m_H}(a,b,\gamma)=&\frac{\pi ^{3/2}  \csc \left(\frac{\pi  \gamma }{2}\right) \Gamma \left(\frac{\gamma }{2}+1\right) \Gamma \left(\frac{\gamma }{2}+2\right)}{2 \left(a^4-1\right)^2 \gamma  \Gamma \left(-\frac{\gamma }{2}\right) \Gamma \left(\frac{\gamma +3}{2}\right)} \Bigg(4 a^{ \gamma +4}-a^{6-\gamma} b^2 (\gamma +2)+a^{2-\gamma }b^2 (\gamma -2) \nonumber\\
&+2 \left(a^4 b^2-2 a^2+b^2\right) a^{ \gamma+2 } \, _2F_1\left(\frac{\gamma }{2},\frac{\gamma +2}{2};\gamma +2;1-a^4\right)\Bigg)+(a\leftrightarrow \frac{1}{a},b\leftrightarrow \frac{1}{b})
\end{align}
In the limit $\gamma\rightarrow0$ the hidden sector reduces to the spurion limit and the formulas simplify drastically. In this limit the model was first discussed in \cite{Dvali:1996cu}, and was later leveraged as a weakly coupled solution to the $A$/$m_H^2$ problem in the special case where $a=b=1$ \cite{Kang:2012ra,Craig:2012xp}. Another interesting special case occurs if $a=1$ and $b=i$ (corresponding to $M_D=M_S$ and $\kappa_D=-\kappa_S$), in which case a symmetry argument forbids both $A_{u,d}$ and $\mu$ at one loop. Both of these special limits serve as important consistency checks of our formulas.  We  elaborate on them further in Appendix \ref{Appspurionlimit}.

\subsection{Solutions to the UV boundary conditions}

As is usual in models with factorizable messenger and hidden sectors, there are some degeneracies in the parametrization of the soft masses in terms of the fundamental parameters of the model. Concretely, all soft masses are left invariant by three different reparametrizations of the fundamental parameters
\begin{align}
& \kappa_T\rightarrow x \kappa_T,\quad M_T\rightarrow x M_T \nonumber \\
&  \kappa_D\rightarrow y \kappa_D,\quad M_D\rightarrow y M_D,\quad \kappa_S\rightarrow y \kappa_S,\quad M_S\rightarrow y M_S,\quad \hat C\rightarrow y^{\gamma} \hat C\nonumber\\
&  M_{D,T}\rightarrow z M_{D,T}, \quad \Lambda \rightarrow z^{\frac{1}{1-\Delta_h}} \Lambda
\label{degneracy1}
\end{align}
where the $x,y$ and $z$ are arbitrary real constants. As we will see in a moment, these degeneracies are relevant when we attempt to map soft parameters onto the various model-specific couplings and mass scales. 

One important subtlety is that the conformal sequestering and the effective OPE coefficient would seem to be degenerate, as can be seen  from (\ref{simpleBmu}), (\ref{simplemH}), and the second line of (\ref{degneracy1}). It would seem to imply that a small effective OPE coefficient with little or no sequestering can be traded for a larger OPE coefficient with more sequestering and vice versa, without affecting the soft parameters. However in practice, the effect of this rescaling is limited by the requirement that the $\kappa_{D,S}$ are perturbative and that $\sqrt{F}<\mathrm{Min}[M_T,M_D,M_S]$.  The two other degeneracies in (\ref{degneracy1}) are restricted by similar consistency conditions.

In general, the model-independent restrictions discussed in section \ref{secparamspace} are supplemented by the additional requirement that the soft parameters can all be realized in terms of the fundamental parameters of the model. In other words, one must establish that there exists a solution to the set of 10 boundary conditions for the soft parameters
\begin{equation}\label{softparams}
 M_{1},M_{2},M_{3}, A_u,A_d,\mu, B_\mu, \mhu^2, \mhd^2\;\mathrm{and}\;\sqrt{F}
\end{equation}
in terms of a realistic choice for the 13 continuous `fundamental' parameters
\begin{equation}
\lambda_{u},\lambda_{d},\kappa_T,\kappa_D,\kappa_S,M_T,M_D,M_S,\hat C,\Delta_h,\gamma,\sqrt{F}\;\mathrm{and}\; \Lambda\label{mapfundtosoft}
\end{equation}
plus the discrete messenger number $N$. Of the 10 soft parameters, only 9 are really independent since we imposed a messenger parity that related $A_d$ to $A_u$, $\mhu$ and $\mhd$. Naively this system of equations appears to be underconstrained, and one would expect that generically a solution should exist. However the situation is bit more subtle. 

First, we have used the results of the conformal bootstrap program (summarized in table \ref{anomalousdim}) to choose the maximum $\gamma$ allowed for a given $\Delta_h$, so they are no longer independent. Secondly, 3 out of the 12 remaining continuous fundamental parameters are degenerate as in (\ref{degneracy1}). For definiteness, we break the degeneracies\footnote{Our choice for $\Lambda$ corresponds to the most optimistic case as far as the impact of the sequestering is concerned. For a different choice of $\Lambda$ the messenger scale and the sequestering can be obtained by the rescaling in (\ref{degneracy1}). } in (\ref{degneracy1}) by fixing $\kappa_T=\kappa_D=2$ and $\Lambda=2\; \mathrm{max}[M_T,M_D,M_S]$. This choice attempts to maximize the impact of the conformal sequestering, while preserving perturbativity in $\kappa_{D,T}$. (Even more sequestering, and thus larger $\hat C$, can be obtained from (\ref{degneracy1}) if one is willing to tolerate a larger value for $\kappa_D$.)

After fixing the degeneracies, we are left with only with 9 independent fundamental parameters to determine 9 independent soft parameters. Since the boundary conditions are highly non-linear in some of the fundamental parameters, a solution is not guaranteed, and requiring its existence can further constrain the acceptable range of the soft parameters in (\ref{softparams}).  Such solutions must be obtained numerically; details on our  algorithm are provided in appendix \ref{Appnumerical}. We do not attempt to find all possible solutions for a given set of soft parameters, but are content with a single viable solution per set of soft parameters.  A `viable' solution in this context means that all masses, couplings and the effective OPE coefficient are real, that the couplings $\lambda_{u}$, $\lambda_d$ and $\kappa_S$ are perturbative and that $\sqrt{F}<\mathrm{Min}[M_T,M_D,M_S]$. The latter will turn out to be a stringent condition if \mbox{$\sqrt{F}\leq10^7$ GeV}.

Table \ref{benchmark} contains an example of a point and its solution in terms of the fundamental parameters for various choices of $\gamma$. Unsurprisingly, conformal sequestering is not efficient for $\gamma=0.1$ and the effective OPE coefficient must be very small to accommodate a solution. For $\gamma=0.4$ and $\gamma=0.7$ on the other hand, conformal sequestering provides roughly an order of magnitude suppression for the one-loop contributions to $B_\mu$ and $\mhud^2$.\footnote{Notice that the sequestering for $\gamma=0.7$ is essentially the same as the sequestering for $\gamma=0.4$, despite the higher anomalous dimension of the former. We have encountered this already in figure \ref{seqcontourplots}. It is due to the competing effects of increased sequestering from larger $\gamma$, but decreased sequestering from larger $\Delta_h$. } 
 Moreover, if we choose $N=6$ the $\frac{1}{N}$ factor in (\ref{simpleBmu}) and (\ref{simplemH}) in combination with conformal sequestering provides a sufficient amount of suppression to facilitate an $\mathcal{O}(1)$ effective OPE coefficient. 

\begin{table}[t!]\centering
\begin{tabular}{|cc||c|c|c|c|}\hline
\multicolumn{2}{|c||}{soft parameters}&\multicolumn{4}{c|}{fundamental parameters}\\\hline
\multicolumn{1}{|c|}{$\sqrt{F}$}&$10^9$ GeV& $\gamma$&0.1&0.4&0.7\\
\multicolumn{1}{|c|}{$M_1$}&1.75 TeV& $\Delta_h$ & 1.20 & 1.45 & 2.00\\
\multicolumn{1}{|c|}{$M_2$}&3.53 TeV& $N$&6&6&6\\
\multicolumn{1}{|c|}{$M_3$}&2.0 TeV& $\lambda_u$&0.66&0.68&0.70\\
\multicolumn{1}{|c|}{$A_u$}&-3.2 TeV& $\lambda_d$&0.60&0.62&0.63\\
\multicolumn{1}{|c|}{$A_d$}&-2.6 TeV& $\kappa_S$&-0.19&-0.25&-0.30\\
\multicolumn{1}{|c|}{$\mu$}&400 GeV& $M_T$&$4.3\times10^{12}$ GeV&$9.0\times10^{11}$ GeV&$1.6\times10^{11}$ GeV\\
\multicolumn{1}{|c|}{$\mhu $}&1.66 TeV& $M_D$&$1.5\times10^{12}$ GeV&$3.2\times10^{11}$ GeV&$4.0\times10^{10}$ GeV\\
\multicolumn{1}{|c|}{$\mhd$}&1.50 TeV&$M_S$&$2.4\times10^{11}$ GeV&$6.4\times10^{10}$ GeV&$9.7\times10^{9}$ GeV\\
\multicolumn{1}{|c|}{$B_\mu$}&0.35 $\mathrm{TeV}^2$&$\left(\frac{\sqrt{F}}{M}\right)^\gamma$&0.53&0.14&0.12\\\cline{1-2}
\multicolumn{2}{c|}{}&$\hat C$&0.071&0.30&0.30\\\cline{3-6}
\end{tabular}
\caption{An example of a point with its interpretation in terms of the fundamental parameters, for various values of $\gamma$. For this point $\tan\beta=10$. \label{benchmark}}
\end{table}

\begin{figure}[p!]\centering
\subfigure[$\gamma=0.1$]{\includegraphics[width=0.45\textwidth]{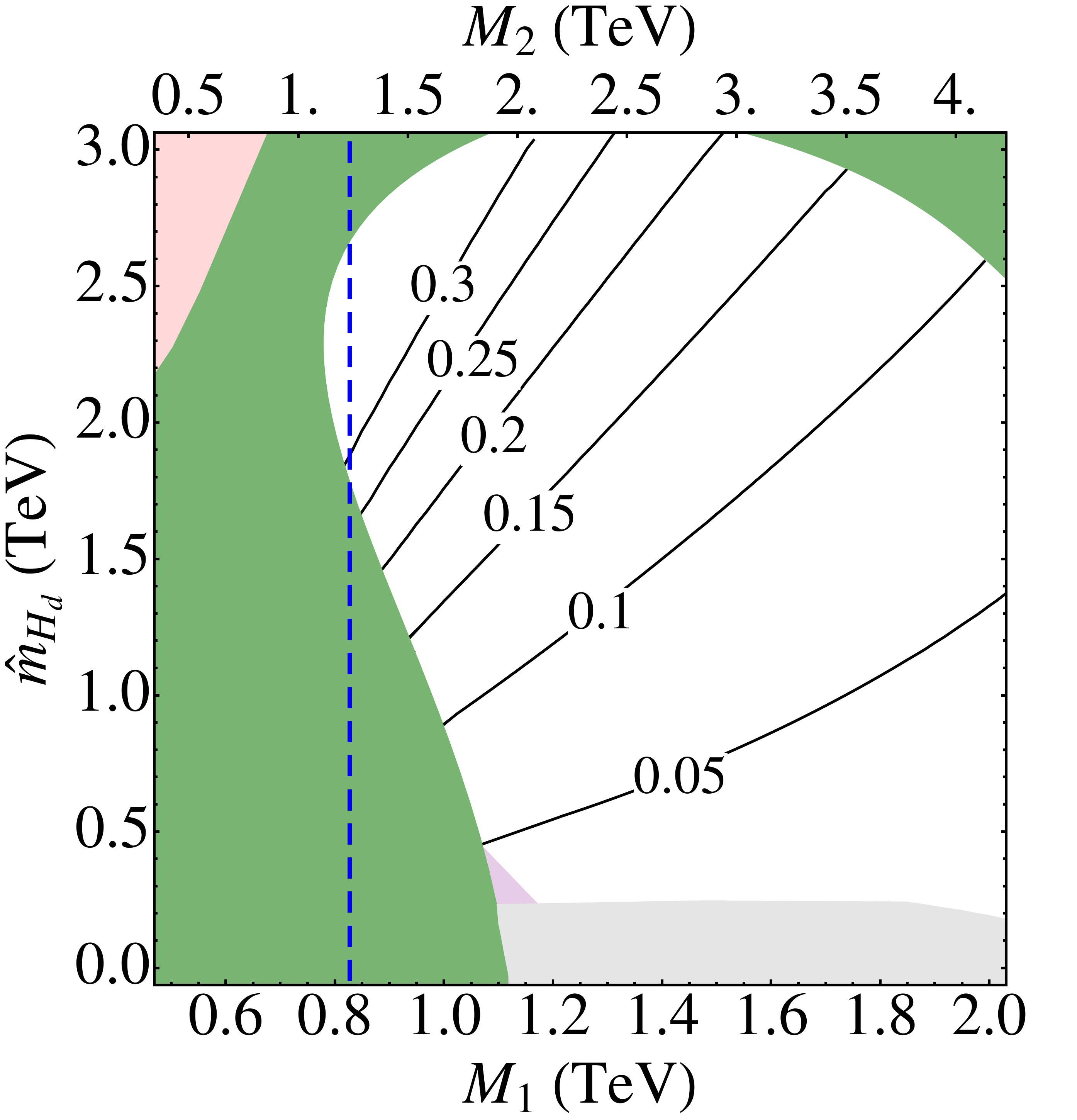}}\hfil
\subfigure[$\gamma=0.4$]{\includegraphics[width=0.45\textwidth]{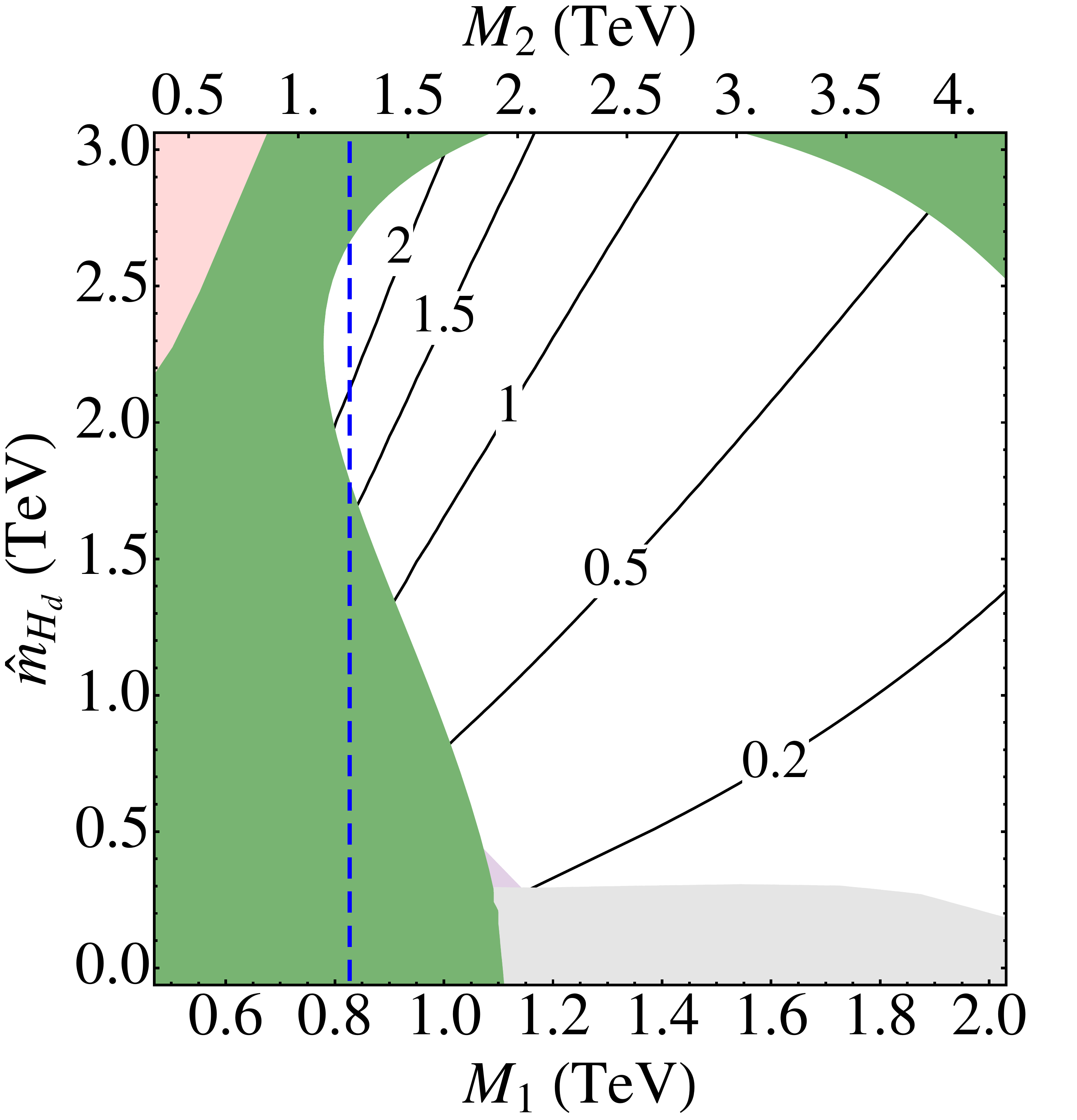}}
\subfigure[$\gamma=0.7$]{\includegraphics[width=0.45\textwidth]{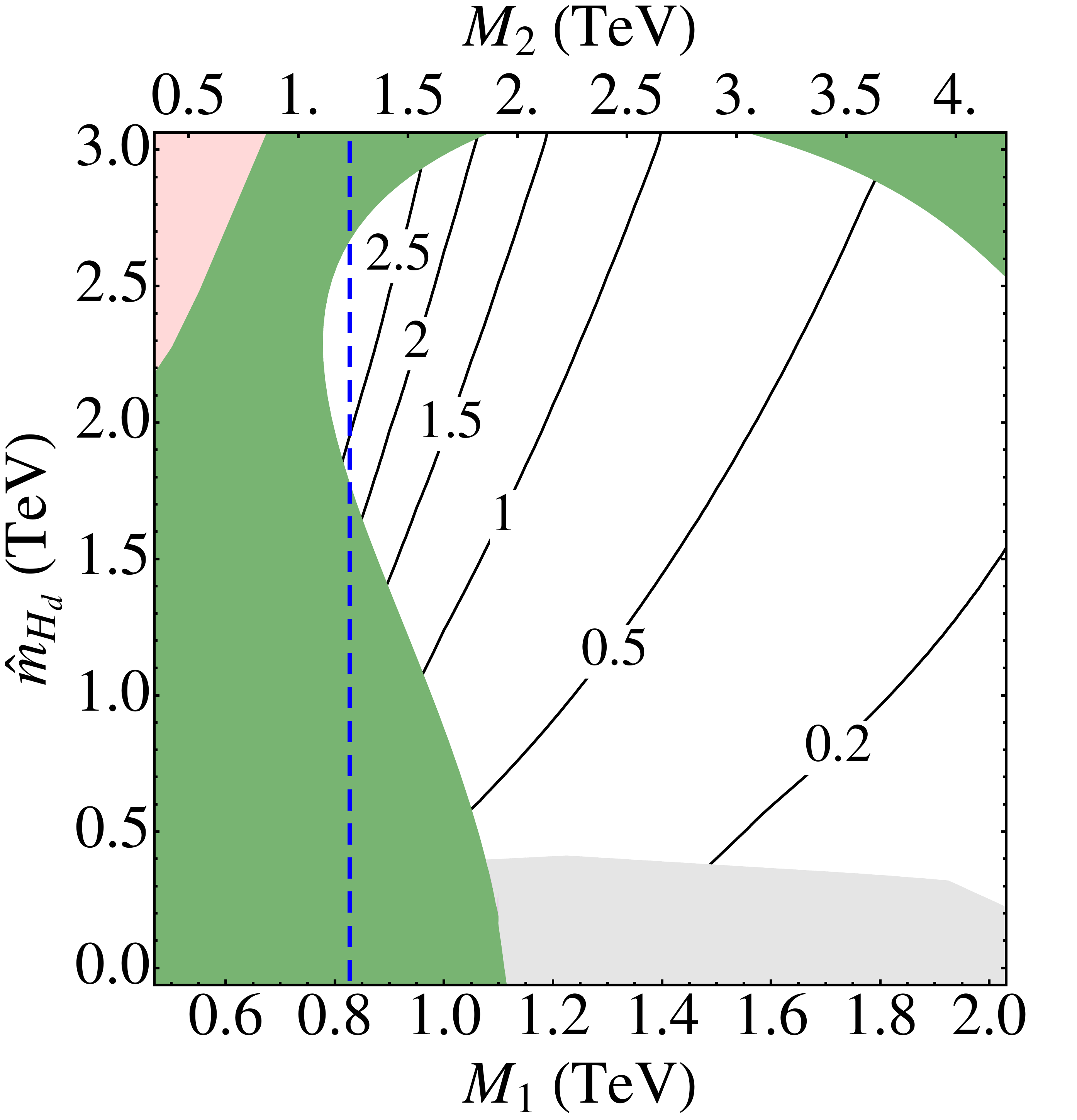}}
\caption{$\hat C$ as a function of  $M_1$ and $\mhd$ for $N=6$. The light gray area indicates the region where our algorithm does not converge on a suitable solution for the UV boundary conditions. All other parameters and colors are as in figure \ref{mastergauginounification}. \label{Ccdependence} }
\end{figure}

More generally, the solutions for the effective OPE coefficient as a function of $M_1$ and $\mhd$ are shown in figure \ref{Ccdependence} for various values of $\gamma$. Almost all of the viable parameter space in figure \ref{mastergauginounification} of the previous section can be covered by our example, except for a small region for low $\mhd^2$ where our numerical method fails to converge on a suitable solution. It is conceivable that these points may be recovered with a more sophisticated numerical procedure. This result suggests that it should be possible to cover the full parameter space with weakly coupled models for the messenger sector; however this is beyond the scope of this work.

\begin{figure}[t!]\centering

\includegraphics[width=0.5\textwidth]{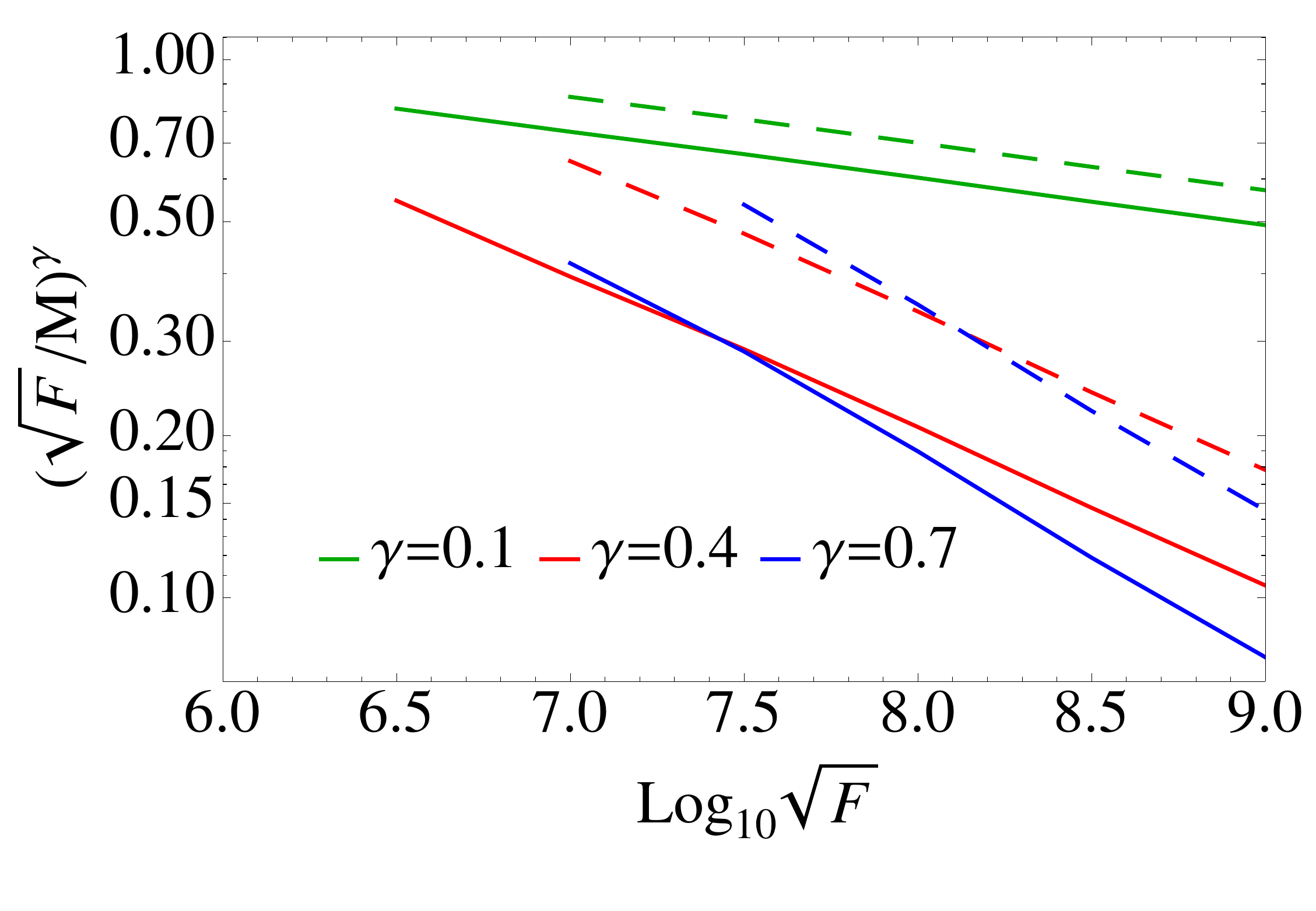}\hfill
\includegraphics[width=0.5\textwidth]{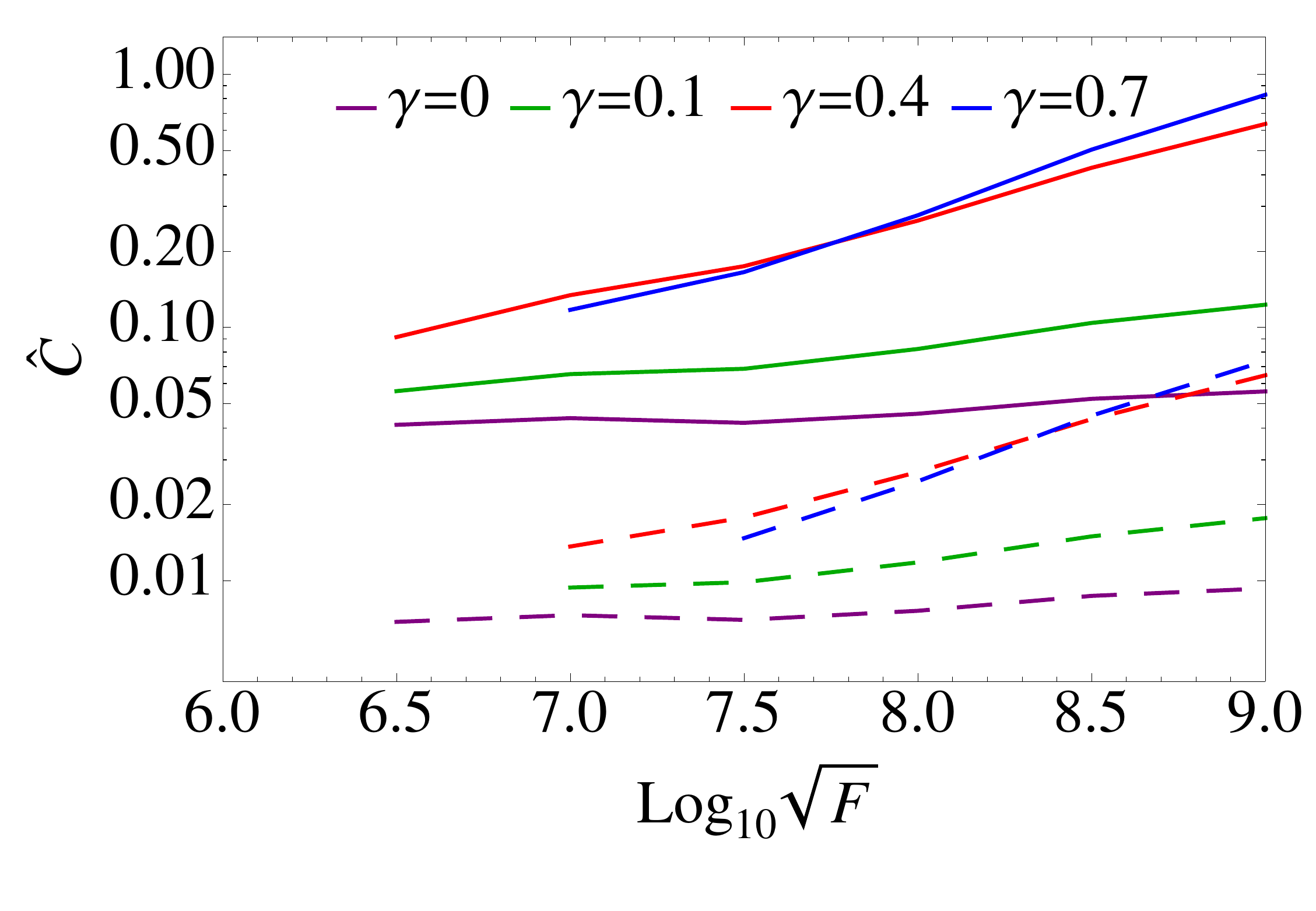}

\caption{Plots of $\left(\frac{\sqrt{F}}{M}\right)^\gamma$ and $\hat C$ as a function of $\log_{10}\sqrt{F}$ for $m_{A^0}=1.5$ TeV, $M_1=1.2$ TeV, $\mu=400$ GeV and $\tan \beta=10$. For each value of $\sqrt{F}$, $M_3$ and $A_u$ are fixed such that $\sqrt{m_{\tilde t_1}m_{\tilde t_2}}$ is minimized under the constraint $m_h>125$ GeV. Dashed lines represent $N=1$, full lines represent $N=6$. The various curves are cut off at the point where the consistency condition $\sqrt{F}<\mathrm{Min}[M_T,M_D,M_S]$ is no longer satisfied. \label{hidsecplot} }

\end{figure}

It is interesting to compare the precise effectiveness of the conformal sequestering in our model as a function of $\sqrt{F}$ with our rough estimates in figure \ref{seqcontourplots}. The sequestering as computed in our example is shown in figure \ref{hidsecplot}, as well as the effective OPE coefficient needed to obtain viable EWSB. In fairly good agreement with our rough estimate in figure \ref{seqcontourplots}, conformal sequestering becomes less efficient for lower $\sqrt{F}$, and its effect completely disappears for $\sqrt{F}\sim 10^6$ GeV. As we have seen, the reason is that for a fixed gaugino mass and lower $\sqrt{F}$, the 
 separation between $M$ and $\sqrt{F}$ must decrease, limiting the capabilities of the sequestering.

 From the left-hand panel of figure  \ref{hidsecplot} one also learns that increasing the messenger number has a double advantage: on the one hand it provides an extra $\frac{1}{N}$ suppression in (\ref{simpleBmu}) and (\ref{simplemH}). In addition, a larger $N$ in (\ref{bino}) allows for a slightly larger splitting between $M_D$ and $\sqrt{F}$ and thus slightly more efficient suppression from conformal sequestering. For low $N$ and low $\sqrt{F}$, a smaller $\hat C$ is needed to compensate for the loss in sequestering and messenger number suppression. This is illustrated in the right-hand panel figure \ref{hidsecplot}, where for completeness we added the extreme limit of $\gamma=0$, which corresponds to no contribution from sequestering. 

\FloatBarrier

\section{Collider Phenomenology\label{colliderphenosec}}
In this section we briefly discuss the collider phenomenology and the current constraints on the model. Since all the sfermion masses are suppressed at the scale $\sqrt{F}$, their IR values are primarily set through gaugino mediation. We emphasize once more that this is a general property of models that attempt to address the $\mu$/$B_\mu$ and the $A$/$m_H^2$ problems with strong hidden sector dynamics. This implies a number of generic features of the low energy spectrum which are independent of the precise content of the hidden and messenger sectors:
\begin{itemize}
\item The gluino tends to be heavier than the squarks, the wino tends to be heavier than the left-handed sleptons, and the bino tends to be heavier than the right-handed sleptons.
\item The colored sfermions are typically heavier than the electroweak sfermions, because only the former are pulled up by $M_3$. One exception is the lightest stop, which may be pushed down due to mixing effects.
\item The NLSP is a stau or a Higgsino and is sufficiently long-lived\footnote{This is assuming R-parity conservation. If R-parity is violated, the NLSP could still decay promptly despite a high supersymmetry breaking scale.}  to escape the detector, except if $\sqrt{F}\sim10^6$ GeV, in which case it decays through a displaced vertex.
\item The LSP is the gravitino if $\sqrt{F}\lesssim 10^{10}$ GeV as we have assumed in this paper. (If $\sqrt{F}> 10^{10}$ GeV, the gravitino mass may be lifted to the extent that it is no longer the LSP \cite{Murayama:2007ge,Craig:2008vs,Craig:2009tz}.)
\end{itemize}

\begin{figure}[t]\centering
\includegraphics[width=0.9\textwidth]{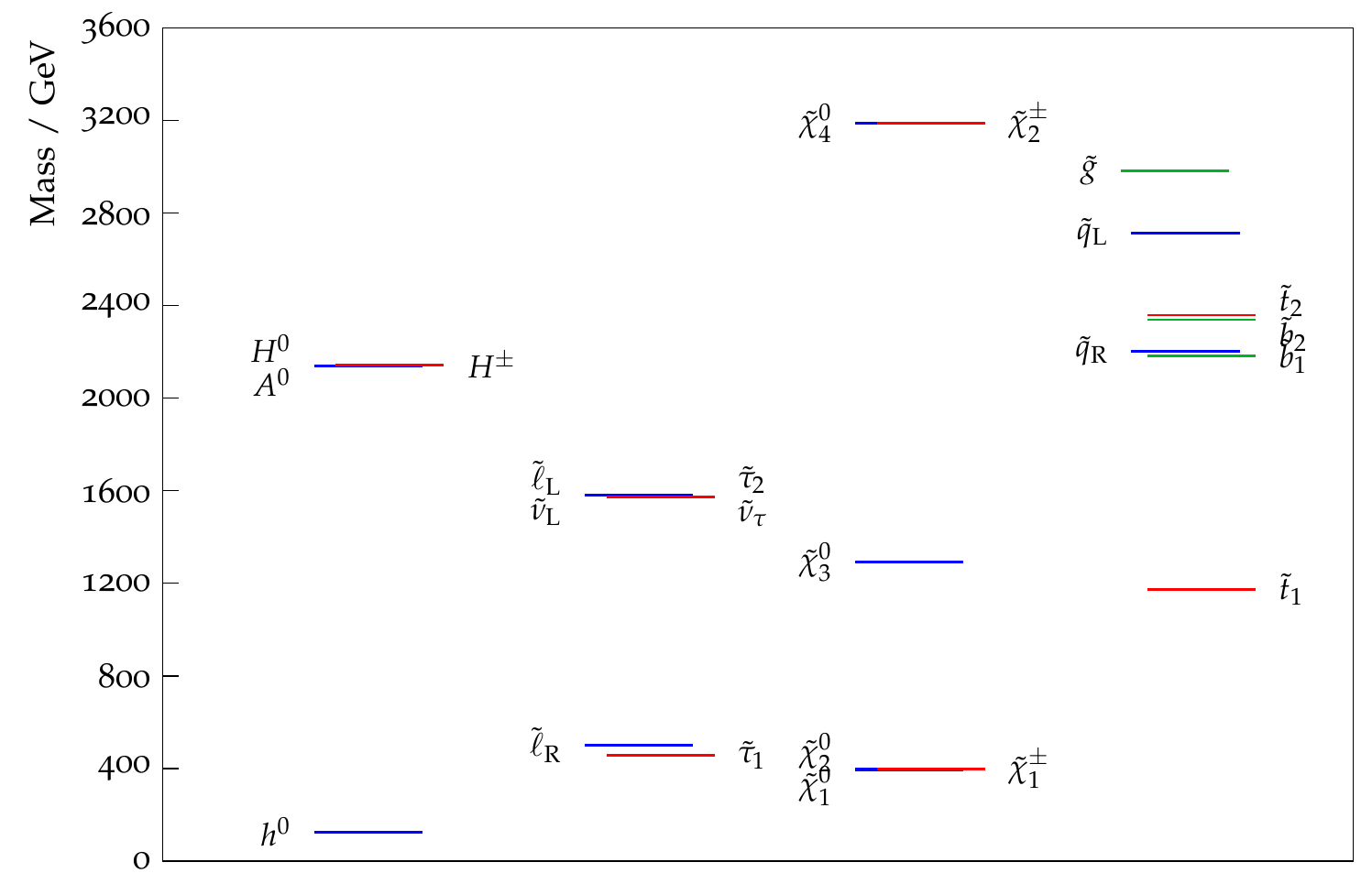}
\caption{The spectrum for the point in table \ref{benchmark}, with Higgsino NLSP. The spectrum with stau NLSP is nearly identical, since the $\mu$ parameter has only a small effect on the other masses.   \label{benchmarkpoint}}
\end{figure}

Unsurprisingly, this class of models is subject to a variety of collider constraints.  Conceptually, it is important to distinguish constraints on the colored part of the spectrum from constraints on the electroweak part. Regarding the former, the masses of the colored sparticles are almost exclusively controlled by $M_3$ and $A_u$. As we saw in section \ref{sec:higgsmass}, these two parameters are determined by the requirement of a $126$ GeV Higgs with TeV-scale stops. Therefore, we expect robust predictions on the typical masses of the colored sparticles. As shown in  figure \ref{moneyplot1}, the lightest stop is always the lightest colored state and must be heavier than 750 GeV, while the minimum gluino mass is roughly 2 TeV.  The masses of the electroweak states on the other hand are controlled by the bino and wino masses, and may be as light as several hundreds of GeV. The phenomenology of these electroweak states is very rich and radically different depending on the nature of the NLSP. In what follows we discuss stau and Higgsino NLSP separately.\footnote{In a narrow corner of the parameter space the sneutrino can be the NLSP. For a discussion on the phenomenology of this scenario we refer to \cite{Katz:2009qx}.} A typical spectrum is shown in figure \ref{benchmarkpoint}.

Since the NLSP is nearly always detector-stable in these models, there are already very powerful collider constraints if the stau is the NLSP. In particular, CMS has excluded such long-lived staus with a cross section above 0.3 fb \cite{Chatrchyan:2013oca}. (A slightly weaker limit from ATLAS is also available \cite{ATLAS-CONF-2013-058}.) This translates to a lower limit on the mass of 339~GeV. As can be seen from figure \ref{general_master_plot}, this constraint implies that the stau NLSP scenario  is now experimentally disfavored.

Since these searches are inclusive, they are also likely to be sensitive to the production of the entire superpartner spectrum, and not just to the staus themselves (see e.g.\ the discussion in \cite{Heisig:2012zq,Heisig:2013rya}). By comparing their production cross sections \cite{susycrosssec} with the CMS limit, one can estimate the bounds on the masses of other sparticles. For instance, we find that the gluino and the stops should be heavier than $\sim 1400$~GeV and $\sim 1000$~GeV respectively. According to the preceding discussion of the colored spectrum (see again  figure \ref{moneyplot1}), this is not a very stringent constraint on these models, where the gluinos and stops are already heavy to begin with. Meanwhile, 
 the Higgsino is excluded below $\sim 600$~GeV, where we estimated the production cross section with Prospino 2.1 \cite{Beenakker:1999xh}. In the discussion in section \ref{secparamspace} we restricted ourselves to $|\mu|<500$ GeV for simplicity, however we verified that Higgsino masses which evade the constraint can easily be obtained.

The collider phenomenology of a detector stable Higgsino NLSP is essentially identical to that of a Higgsino LSP in gravity mediation models. Unsurprisingly, from an experimental point of view a long-lived Higgsino NLSP is much more challenging than a long-lived stau NLSP. If the other states decouple, the only robust bound comes from LEP, and requires the charged Higgsino component to be heavier than $92.4$ GeV \cite{LEPHiggsino}. Since the sfermion masses are generated through gaugino mediation, the next state in the spectrum is typically the lightest stau mass eigenstate, possibly degenerate with the right-handed light flavor sleptons. With a Higgsino heavier than the LEP bound, there is currently no bound on these right-handed sleptons if they are Drell-Yan produced \cite{ATLAS-CONF-2013-049,CMS-PAS-SUS-13-006}. The left-handed sleptons on the other hand have a higher production cross section and are constrained to be heavier than 300 GeV if the Higgsino is lighter than 160 GeV \cite{ATLAS-CONF-2013-049,CMS-PAS-SUS-13-006}, however this bound does not yet significantly constrain our minimal model with  ${\bf 5}$-${\bf \bar 5}$ messengers. The lightest colored state is always the lightest stop and is always outside the reach of the 8 TeV LHC, but its direct production could be a promising channel at the 14 TeV run. Although the spectrum is not natural in the strict sense, this signature is covered by existing ``Natural SUSY" search strategies. 
\FloatBarrier

\section{Discussion and Outlook}

Strong hidden sector dynamics may provide an elegant framework in which both the $\mu$/$B_\mu$ and the $A$/$m_H^2$ problems can be addressed through a single mechanism. Rather than relying on a cleverly designed messenger sector, this class of models counters the disastrous $16\pi^2$ enhancement of $B_\mu$ and $m_H^2$  by a suppression from strong dynamics in the hidden sector. This suppression can arise from conformal sequestering, a small effective OPE coefficient, large messenger number, or a combination of all three. We provide a simple example of a complete model, as well as the first explicit calculation of the low energy observables in terms of scaling dimensions, vacuum expectation values and OPE coefficients of the leading operators in the hidden sector. The essential tool enabling this calculation is the GMHM framework \cite{Craig:2013wga}. 

Accounting for the bounds on the anomalous dimension from the conformal bootstrap program \cite{Poland:2011ey}, we make a general estimate of the impact of conformal sequestering for this class of models and validate our estimate in an explicit example. In either case, conformal sequestering is insufficient to produce a full loop factor suppression, but a suppression of roughly one order of magnitude is possible if $\sqrt{F}\sim10^9$ GeV. In this case viable electroweak symmetry breaking can be achieved for effective OPE coefficients roughly between 1 and 0.1, depending on the details of the messenger sector. It is still an open question whether the upper bound on $\gamma$ from the bootstrap program can be saturated, as currently no examples are known. Such an example would necessarily need to be strongly coupled, as weakly coupled SCFT's were shown not to produce the required inequalities for the scaling dimensions of the operators  \cite{Green:2012nqa}.

An important and generic feature of this class of models is that the suppression from conformal sequestering is only appreciable for $\sqrt{F}$ as high as roughly $10^9$ GeV. From this fairly high scale of supersymmetry breaking one would expect a degree of fine-tuning of roughly 1 in $10^3$ at best, and a priori the tuning may be aggravated by large cancellations in the UV boundary conditions, as is the case in Higgs mediation models without appreciable hidden sector dynamics. In most of the parameter space of our example such large cancellations do not occur, indicating that the tuning estimates from the high scale RG running are a fair estimate of the total tuning of the model. The model therefore constitutes a solution to the ``little $A$/$m_H^2$ problem'' as presented in \cite{Craig:2012xp}, although a moderate price in tuning had to be paid from the higher supersymmetry breaking scale.

This problem would be alleviated to some degree by considering lower values $\sqrt{F}$, where the suppression of the loop factor must be obtained from the smallness of the effective OPE coefficient rather than from the conformal sequestering.\footnote{Alternatively, we could conceivably avoid the loop factors altogether with a strongly-coupled messenger sector. Although such a setup may greatly alleviate the fine-tuning by allowing for lower $\sqrt{F}$, it may also lose much of its predictivity and calculability. 
}  At this point it is not clear whether such small OPE coefficients can be achieved in a realistic model. The conformal bootstrap program has resulted in interesting lower bounds on OPE coefficients, provided that there is a gap in the spectrum of operators \cite{Poland:2011ey}. Implicitly, we assumed the existence of such a gap by truncating the OPE after the leading term, and it seems plausible that our scenario may be constrained from this end as well.\footnote{We thank David Simmons-Duffin for bringing this to our attention.} A detailed quantitative analysis of this type of constraint is beyond the scope of this paper, but is certainly worth exploring.

Even if very small OPE coefficients could be made compatible with the bootstrap constraints, within our simple example, we found it very challenging to find viable solutions with $\sqrt{F}\sim10^6$ GeV. But we strongly suspect that even extending the model slightly would allow for many more solutions with low $\sqrt{F}$. A broader question which is also interesting is whether it is possible to completely cover the rest of the parameter space in (\ref{paramspace}). It is encouraging that even with our simple example we were able to sample a large part of it. We therefore suspect that it should be possible to cover the full parameter with a set of perturbative messenger models. Here are some promising ideas in this direction. First we could relax our assumption on the action of the messenger parity on $O_u$ and $O_d$. For instance, one could consider multiple portals between the messenger sector and the MSSM Higgs sector, of the form:
\begin{align}
W\supset \sum_i\lambda^{(i)}_u O^{(i)}_u H_u+\sum_i\lambda^{(i)}_d O^{(i)}_d H_d\label{extensions}
\end{align}
Another idea would be to allow for   portals of the form
\begin{align}
W\supset \lambda S H_uH_d\
\end{align}
where $S$ is a gauge singlet. This is interesting since the singlet portal does not generate $A_{u,d}$ and $\mhud^2$ at the same loop order as $\mu$ and $B_\mu$ \cite{Komargodski:2008ax} and therefore provides a clean way to untangle these two soft parameters from the others. Finally, in our simple model we assumed ${\bf 5}$-${\bf\bar 5}$ messengers. A model including ${\bf 10}$-${\bf \overline{10}}$ messengers (as in \cite{Asano:2008qc}) would offer more parametric freedom.

In our analysis we restricted ourselves to $\sqrt{F}<10^{10}$ GeV in order to avoid problems with charged LSPs and to automatically eliminate Planck-induced flavor violation. However it is  conceivable that with enough assumptions about the hidden sector, conformal sequestering could also suppress dangerous Planck-induced operators \cite{Randall:1998uk,Luty:2001jh,Luty:2001zv}. If this is true, then our model could be extended beyond $\sqrt{F}\sim 10^{10}$ GeV (at least with the Higgsino being the LSP). Such a scenario deserves further study, especially since with larger $\sqrt{F}$, the impact of the conformal sequestering can be further enhanced beyond what we have found in this paper.

The collider phenomenology in this class of models is generically similar to the phenomenology of gaugino mediation with large $A$-terms and depends strongly on the nature of the NLSP. If the NLSP is the Higgsino, the phenomenology is similar to that of a neutralino LSP. The constraints on this scenario are currently rather weak and prospects for the 14 TeV run depend heavily on the spectrum of the colored states. On the other hand, if the NLSP is a stau, our model is already strongly constrained by current searches. Moreover, in this case direct stop production would be a spectacular channel at the 14 TeV run of the LHC, which should allow us to definitively test this scenario.

 \vspace{0.5cm}

\noindent\textbf{Acknowledgements}
We thank Nathaniel Craig for fruitful collaboration in the early stages of this work and useful discussions later on. We further thank  Surjeet Akula, Patrick Draper, Jan Heisig, Satoshi Shirai and David Simmons-Duffin for useful discussions. Moreover we are grateful to Ben Allanach and Ben O'Leary for their advice regarding \verb+SOFTSUSY+ and \verb+Vevacious+ respectively. The work of DS is supported in part by a DOE Early Career Award (DOE-ARRA-SC0003883) and an Alfred P.~Sloan Foundation Fellowship.  The work of SK is supported in part by DOE grants DOE-SC0010008, DOE-ARRA-SC0003883 and DOE-DE-SC0007897. 

\appendix

\section{Simplifying Limits\label{Appspurionlimit}}
The boundary conditions for $B_\mu$ and $\mhud$ for our example in section \ref{example} simplify dramatically in the spurion limit ($\gamma\rightarrow0$ and $\hat C\rightarrow 1$). Specifically, the dimensionless functions reduce to:
\begin{align}
f_B(a,b,0)=&\frac{1}{\left(a^4-1\right)^3}\Big[{b^2 \left(-a^8+8 a^4 \log a+1\right)}\\&
-{a^2 \left(a^4+1\right) \left(1-a^4+2 \left(a^4+1\right) \log a\right)}\Big]+(a\leftrightarrow 1/a,b\leftrightarrow 1/b )\nonumber\\
f_{m_H}(a,b,0)=&-\frac{2 a^2 (a^2-b^2) }{\left(a^4-1\right)^3}\left(1-a^4+2 \left(a^4+1\right) \log a\right)+(a\leftrightarrow 1/a,b\leftrightarrow 1/b )\label{labelmhspurion}
\end{align}
If in addition we take $a=b$, the model reduces to the model first presented by Dvali, Giudice and Pomarol \cite{Dvali:1996cu} and we can verify that in this limit our results agree with theirs. Concretely, the dimensionless functions further reduce to
\begin{align}
f_\mu(a,a)&= \frac{a^2\log a^4}{1-a^4}\nonumber\\
f_A(a,a)&=-1 \nonumber\\
f_B(a,a,0)&=\frac{a^2\log a^4}{1-a^4} \nonumber\\
f_{m_H}(a,a,0)&=0
\end{align}
Observe that $\mhu$ and $\mhd$ vanish at one loop; this was the basis of the weakly coupled solution to the $A$/$m_H^2$ problem presented in \cite{Craig:2012xp}. In that paper we considered the special limit $\lambda_d=0$ which ensures that $\mu$, $B_\mu$, $A_d$ and $\mhd$ vanish. In such a setup, the $\mu$/$B_\mu$ problem is postponed and must dealt with separately, for instance by extending the MSSM with an extra singlet.

For the case $a\ne b$, DGP also provide an expression for $\mhud^2$.\footnote{To 1 loop order, the distinction between $\mhud^2\equiv m^2_{H_{u,d}}+|\mu|^2$ and $m^2_{H_{u,d}}$ is moot. } Their notation is somewhat different from ours, and the $\Lambda_1$ and $\Lambda_2$ in equation (22) of \cite{Dvali:1996cu} correspond to 
\begin{align}
\Lambda_1=\frac{a}{b}\Lambda_H \quad \mathrm{and}\quad \Lambda_2=\frac{b}{a}\Lambda_H
\end{align}
With this change of notation in (\ref{labelmhspurion}), our expression for $\mhud^2$ becomes
\begin{align}
\mhud^2=&\frac{|\lambda_{u,d}|^2}{16\pi^2}(\Lambda_1-\Lambda_2)^2 g(a) 
\end{align}
with
\begin{align}
g(a)=-a^4\frac{1-a^4+(1+ a^4) \log (a^4)}{\left(1-a^4\right)^3}
\end{align}
The magnitude of our expression agrees with equation (22) in \cite{Dvali:1996cu}, however we disagree on the sign. We find $\mhud^2>0$ and since the Higgs fields can be considered as pseudomoduli in a model with only fields of R-charge 0 and 2, we have confidence in our result \cite{Shih:2007av,Curtin:2012yu}.

The second interesting special limit is when $a=1$ and $b=i$, as in this case $\mu$ and $A_{u,d}$ vanish at one loop. The superpotential reduces now to 
\begin{align}
W=\kappa\frac{{O}_h}{\Lambda^{\Delta_h-1}}\left(\tilde\phi_D\phi_D- \tilde\phi_S\phi_S\right)+M\left(\tilde\phi_D\phi_D+ \tilde\phi_S\phi_S\right)+\lambda_{u}\tilde\phi_D\phi_S H_u+\lambda_{d} \phi_D\tilde\phi_S H_d.\label{Wspecialcase}
\end{align}
with $\kappa\equiv\kappa_D=-\kappa_S$ and $M\equiv M_D=M_S$. The model now has an enhanced discrete symmetry: 
\begin{align}
\phi_D\leftrightarrow \tilde\phi_S\quad \tilde\phi_D\leftrightarrow \phi_S \quad{O}_h\rightarrow -{O}_h\label{discretesym}
\end{align}
which forbids the correlators (\ref{eq:mu}) and (\ref{eq:A}) at the one loop level\footnote{Of course the discrete symmetry does not commute with the gauge symmetry, and is therefore not a symmetry of the full theory. However for the 1 loop Higgs mediated contributions the gauge charge of $\phi_D$ is irrelevant.} since the operator $O_m$ is odd under (\ref{discretesym}). This feature may be useful when attempting to cover the full GMHM parameter space with weakly coupled models for the messenger sector.

\section{Numerical Procedure\label{Appnumerical}}
Our general philosophy is to front-load the part of the calculation that involves integrating the RGE's, and delay the implementation of the model-specific boundary conditions as long as possible. This allows us to study various tachyons and EWSB requirements in terms of the familiar soft parameters, rather than the somewhat unintuitive parameters $\lambda_{u,d}$, $\hat C$ etc. This approach also should allow for a more straightforward generalization to other models, since the model-independent, more time consuming steps are performed first. Concretely, we parametrize our scan in terms of the independent variables
\begin{equation}\label{params2}
M_{1,2,3}, A_u,A_d,\tan\beta, \mu,m_A(\mathrm{pole})\; \mathrm{and}\;\sqrt{F} 
\end{equation}
where all parameters are specified at the scale $\sqrt{F}$, except the pole mass of the pseudoscalar $m_A$. We choose the latter rather than $\mhd$ such that our scan is maximally compatible with the inputs that must be provided to \verb+SOFTSUSY-3.3.9+ \cite{Allanach:2001kg}. For the case where the messengers fit into ${\bf 5}$-${\bf \bar 5}$  representations, we solve for $M_2$ from the outset by using (\ref{gauginorelation}). Our method can be further broken down in the following steps:

\begin{enumerate}
\item  For a given choice of (\ref{params2}), \verb+SOFTSUSY-3.3.9+ computes the RG-running and imposes the EWSB conditions, a procedure which results in a value for $\mhu^2$ and $\mhd^2$ at the scale $\sqrt{F}$. Furthermore we determine $A_d$ as a function of the other soft parameters by imposing (\ref{Adrelation}). Since (\ref{Adrelation}) involves $\mhu^2$ and $\mhd^2$, this must be done through an iterative procedure, which we repeat until convergence is achieved. At this point we are done with integrating the RGE's, and there is no more need to run \verb+SOFTSUSY-3.3.9+ in the remainder of the calculation.

\item At this stage we can express $\Lambda_H$ as a function of $M_2$, $a$ and $b$. At the level of finding a solution for the boundary conditions, the variables $\hat C$, $N$ and $\left(\frac{\sqrt{F}}{M}\right)^\gamma$ are degenerate. We therefore define an auxiliary variable
\begin{align}
\tilde C \equiv \frac{\hat C}{N}\left(\frac{\sqrt{F}}{M}\right)^\gamma\label{Ctilde}
\end{align}
to simplify the solution finding procedure. Solving the UV boundary conditions specified in (\ref{simplemu}), (\ref{simpleA}), (\ref{simpleBmu}) and (\ref{simplemH}) thus corresponds to solving 6 algebraic equations in terms of the 5 variables $\lambda_{u,d}$, $a$, $b$ and $\tilde C$. In the previous step we already eliminated $A_d$ by solving (\ref{Adrelation}) through the iterative procedure. This leaves us with 5 equations with 5 unknowns, and a much better chance of obtaining a viable solution than if we would have attacked all 6 equations at once. This translates into a much improved computation time per point than if we would have performed a brute force scan over $\lambda_{u,d}$, $a$, $b$ and $\tilde C$. Next we can isolate a simple set of two equations by taking a clever combination of the boundary conditions:
\begin{align}
\frac{\mu^2}{A_u A_d}&=\left(\frac{f_\mu(a,b)}{f_A(a,b)}\right)^2\nonumber\\
\frac{B_\mu^2}{(\mu^2+m^2_{H_u})(\mu^2+m^2_{H_d})}&=\left(\frac{f_B(a,b,\gamma)}{f_{m_H}(a,b,\gamma)}\right)^2
\end{align}
This system of equations is independent of $\lambda_{u,d}$ and $\tilde C$ and can be solved analytically for $b$. At this point we have to commit to a concrete choice of $\gamma$, after which we can solve the remaining equation for $a$ numerically.

\item Now that we have solved for $a$ and $b$ for a given choice of $\gamma$, it is trivial to solve the remaining boundary conditions for $\lambda_{u,d}$ and $\tilde C$. At this point we discard the solution if any of these parameters does not have a real solution, if $|\lambda_{u}|>3$, if $|\lambda_{d}|>3$ or if $\tilde C > 100$. These cuts are chosen arbitrarily to ensure no non-physical solutions where kept. We verified that the results are not sensitive to the precise value of these cuts.

\item In the final step we recover $M_D$ from (\ref{bino}) and table \ref{anomalousdim}, and use this to unpack $\tilde C$ in terms of the suppression factor from conformal sequestering and the effective OPE coefficient $\hat C$. At this step we also must commit to a choice of messenger number $N$. 

\end{enumerate}

By delaying an explicit choice for $\gamma$ and $N$ as long as possible we gained in both flexibility and computation speed. 

\providecommand{\href}[2]{#2}\begingroup\raggedright\endgroup

\end{document}